\documentclass[prd,aps,preprint,preprintnumbers,nofootinbib]{revtex4}
\usepackage{epsfig,footnote}
\usepackage{ulem}
\usepackage{color}
\usepackage{array}
\usepackage{amssymb}
\usepackage{amsmath}
\usepackage{graphicx,subfigure}
\usepackage{longtable}
\usepackage{verbatim}
\usepackage{amsfonts}
\usepackage{hyperref}
\usepackage{cancel}

\begin{document}
\title{Analysis of three-body hadronic $D$-meson decays}
\author{Xu-Da Zhou and Si-Hong Zhou\footnote{corresponding author: shzhou@imu.edu.cn}}

\affiliation{Inner Mongolia Key Laboratory of Microscale Physics and Atom Innovation, 
School of Physical Science and Technology, 
Inner Mongolia University, Hohhot 010021, China}
\affiliation{Center for Quantum Physics and Technologies, 
School of Physical Science and Technology, 
Inner Mongolia University, Hohhot 010021, China}

\date{\today}
\begin{abstract}
Motivated by recent experimental advances in  three-body hadronic 
$D$ decays from BESIII, we present a systematic analysis of 
$D_{(s)} \to P_1 (V \to)  P_2 P_3 $ decay processes, where 
$V$ denotes vector resonances ($\rho, K^*, \omega$, $\phi$) 
and $P_{1,2,3}$ are light pseudoscalar mesons ($\pi,K, \eta^{(\prime)}$). 
Using the factorization-assisted topological-amplitude (FAT) approach,
we calculate the intermediate subprocesses $D_{(s)} \to P_1 V$,
incorporating relativistic Breit-Wigner distributions to model the 
subsequent $V \to P_2 P_3$ strong decays. By comprehensively 
including all relevant resonances ($\rho,  K^*, \omega, \phi$), 
we calculate branching fractions for these decay modes 
as well as the Breit-Wigner-tail effects in $D_{(s)} \to P_1 (\omega \to) KK$ 
processes. Our framework comprehensively incorporates both 
factorizable and nonfactorizable contributions, significantly improving 
theoretical predictions in the nonperturbative regime where conventional 
methods face challenges due to the limited mass scale of charm mesons.
The FAT approach yields results in good agreement with experimental data,
 demonstrating its effectiveness in capturing nonfactorizable contributions 
 with improved precision. Our predictions for yet-unobserved decay modes,
particularly those with branching fractions in the order of $10^{-4}\sim10^{-3}$, 
are expected to be tested in future high-precision experiments at BESIII and LHCb.

\end{abstract}

\maketitle

\section{Introduction}\label{Introduction}
Hadronic charm decays are important for probing  the dynamics of 
both strong and weak interactions in the low-energy regime.
Particularly in multibody $D$-meson decays, such as three-body modes, 
with nontrivial kinematics and phase space distributions,
they offer valuable opportunities for exploring hadron spectroscopy 
as the invariant mass squared of two final-state particles often peaks, 
revealing resonances at the edges of the Dalitz plot.

Experimentally, measurements of the branching fractions and {\it CP} asymmetries
of hadronic $D$-meson decays, by BESIII~\cite{BESIII:2016nrs,BESIII:2018hui,
BESIII:2018exp,BESIII:2018pku,BESIII:2019xhl,BESIII:2021jnf}, 
LHCb~\cite{LHCb:2011nqf,LHCb:2013qzm,LHCb:2014nnj, LHCb:2023qne,LHCb:2024rkp},
CLEO~\cite{CLEO:2000fvk,CLEO:2007rrw,CLEO:2008msk,CLEO:2010enr,CLEO:2011cnt,CLEO:2012obf, CLEO:2013bae,CLEO:2013rjc}, Belle (II)~\cite{Belle:2008qrk,Belle-II:2024gfc}, 
and BABAR~\cite{BaBar:2005rek,BaBar:2006dxf,BaBar:2008xzl,BaBar:2008nlp,BaBar:2012ize},
offer direct insights into the amplitudes and phases 
involved in the decay process. 
For multibody $D$-meson decays, in addition to the absolute branching fractions
obtained in Refs.~\cite{ BaBar:2006dxf, CLEO:2007rrw, BESIII:2016nrs, BESIII:2018hui,BESIII:2018exp,BESIII:2018pku,BESIII:2019xhl,BESIII:2021jnf}, 
the fit fractions of each resonance and nonresonance components
can also be measured. This can be done using the amplitude analysis with the Dalitz plot technique 
where the resonances are parametrized as the relativistic Breit-Wigner (RBW) model
and nonresonant terms by exponential distributions.
Especially, three-body $D$-meson decays were predicted to occur primarily 
through intermediate quasi-two-body states in Ref.~\cite{Buras:1985xv}, 
which aligns with the experimentally observed pattern~\cite{BaBar:2005rek,BaBar:2007soq,BaBar:2008nlp}.
The experimental measurements of these quasi-two-body decays 
provide new information on the resonances.

 On the theoretical side, 
given that the charm quark mass $m_c$ is not heavy enough and $1/m_c$ power corrections 
are so large that a reliable heavy quark expansion becomes impractical,
 the QCD-inspired approaches, such as QCD factorization (QCDF)~\cite{Beneke:2000ry}
 and (perturbative QCD) PQCD~\cite{Lu:2000em,Keum:2000wi} 
 applied successfully in hadronic $B$ decays might not be suitable for describing
 hadronic decays of $D$ meson.
 The studies of $D$-meson decays involve application of approximate flavor symmetries, 
such as flavor $SU(3)_F$~\cite{Ryd:2009uf, Bhattacharya:2008ke,Wang:2021ail} 
for three-body decays, topological diagrams for two-body modes~\cite{Brod:2011re,Cheng:2010ry, Cheng:2012wr,Cheng:2012xb,Chen:2012am,Cheng:2016ejf,Cheng:2019ggx,Hsiao:2019ait}
and its implication in three-body decays~\cite{Cheng:2021yrn}, 
the factorization-assisted topological-amplitude (FAT) approach for two-
~\cite{Li:2012cfa,Li:2013xsa,Qin:2021tve} and 
three-body decays~\cite{Zhou:2018suj, Wang:2025rkr},
SU(2)-subgroup symmetry, and U spin~\cite{Dery:2021mll},
and a series of works on scalar resonance in three-body $D$-meson
 decays using flavor $SU(3)_F$~\cite{Molina:2019udw,Toledo:2020zxj, Roca:2020lyi,Dai:2021owu,Dai:2023jix,Bayar:2023azy,Ikeno:2024fjr,Ikeno:2021kzf}.
A phenomenological approach based on QCDF for 
three-body hadronic $D$-meson decay amplitudes can be found in 
Refs.~\cite{Boito:2017jav, Guo:2018orw,Yu:2021euw,Song:2025lmj}, while
the approaches using more complex theoretical background 
in Faddeev techniques and Khuri-Treiman and triangle singularities are discussed in
~\cite{Niecknig:2015ija,Nakamura:2015qga,Magalhaes:2011sh,Aoude:2018zty},
and an inverse problem approach is discussed in~\cite{Li:2020xrz,Saur:2020rgd}.
As mentioned at the beginning, since the low-energy scale involved 
in $D$-meson decay processes, 
the theoretical studies on $D$-meson decays remain relatively limited.

 In this work, we focus on three-body $D$-meson decays 
 $D_{(s)} \to  P_1(V \to) P_2 P_3$, where $V$ represents a vector resonant, $\rho, K^*, \omega, \phi$,
 and $P_{1,2,3}$ are light pseudoscalar mesons, specifically pion, kaon and $\eta^{(\prime)}$.  
 These decays are often referred to quasi-two-body decays, as two of the three final-state particles
 $P_2$ and  $P_3$ originate from an intermediate vector resonance $V$, while the third meson $P_1$,
 referred to as the ``bachelor" meson, recoils against the resonance system.
 Such processes typically manifest at the edges of the Dalitz plot, and
 their amplitude analyses are accessible in experiments.
 BESIII Collaborations have performed the amplitude analysis and 
 measured the partial branching fractions of quasi-two-body $D$ decays, 
 such as  $D^+ \to K_S^0 \pi^+ \pi^-$, $D^+ \to K_S^0 \pi^+ \pi^0$,
 and $D_s^+ \to K^+ \pi^+ \pi^- $ with $\rho$ and $\bar K^*$ as 
 intermediate resonances~\cite{BESIII:2014oag,BESIII:2022vaf},
$D^0 \to K_S^0 (\phi \to) K^+ K^-$~\cite{BESIII:2020hfw},
$D_s^+ \to \pi^+ (K^*, \phi \to) K^+ K^-$~\cite{BESIII:2020ctr},
$D_s^+ \to K_S^0 (\rho, K^* \to) \pi^+ \pi^0$~\cite{BESIII:2021xox}, 
$D^+ \to K_S^0 (K^* \to ) K^+ \pi^0$~\cite{BESIII:2021dmo},
and doubly Cabibbo-suppressed decays 
$ D^+ \to \pi^0 / \eta (K^* \to ) K^+ \pi^0$~\cite{BESIII:2021qsa},
$D_s^+ \to \eta^\prime (\rho \to) \pi^+ \pi^0$~\cite{BESIII:2022ewq}.
 LHCb Collaborations have investigated structures of vector resonance $\bar {K}^{*0}, \phi $
 with their corresponding fit fractions through $D^+ \to K^- K^+ \pi^+$~\cite{LHCb:2011nqf}.
 This experimental information on $D_{(s)} \to  P_1(V \to) P_2 P_3$ 
 can be used to explore the resonant substructure and understand strong dynamics of final particles.
  
 For the systematic study of $D_{(s)} \to  P_1(V \to) P_2 P_3$ decays, 
 we will apply the FAT approach. 
 It was initially proposed just to address the issue of nonfactorizable contributions 
in two-body $D$ and $B$ meson decays by one of us
(S.-H. Zhou and coworkers~\cite{Li:2012cfa, Li:2013xsa, Zhou:2015jba, Zhou:2016jkv, 
 Jiang:2017zwr, Wang:2017ksn, Zhou:2019crd, Zhou:2021yys,Qin:2021tve}), and 
we have subsequently extended it to describe quasi-two-body $B$ 
meson decays~\cite{Zhou:2021yys,Zhou:2023lbc,Zhou:2024qmm}.
This theoretical framework is built upon the conventional topological 
diagram approach~\cite{Cheng:2010ry},
which categorizes decay amplitudes according to distinct electroweak Feynman diagrams 
while retaining SU(3) breaking effects.
It incorporates SU(3) asymmetries into topological diagram amplitudes 
primarily through form factors and decay constants, assisted by QCD factorization. 
The remaining contributions are treated as a small number of unknown 
nonfactorization parameters, which are determined through a global fit to 
all available experimental data.
 Thanks to recent measurements of various $D \to PV$ decay modes
 with a significantly improved precision measured by BESIII, 
 particularly with the addition of some single and doubly Cabibbo-suppressed modes, 
 we have updated the analysis of $D \to PV$  decays in the FAT approach~\cite{Zhou:2025}.
 By incorporating these new and precise data of Cabibbo-favored, 
 single Cabibbo-suppressed, and doubly Cabibbo-suppressed modes, 
 we have refitted the nonfactorizable parameters. 
 This global fit analysis differs from the previous results obtained 
 using the topological diagram approach, which was limited to Cabibbo-favored mode data 
 and led to many solutions of the topological amplitude parameters 
 with similarly small local minima in $\chi^2$ in Ref.~\cite{Cheng:2024hdo}. 
 Consequently, the FAT approach now provides the most accurate decay amplitudes for 
 two-body $D$ meson decays. We will use the updated nonfactorizable parameters 
 to calculate the branching fractions of quasi-two-body $D$ decays, while 
leave the issue of {\it CP} asymmetries arising from interference 
between tree and penguin processes for the future work
owing to the lack of knowledge on precise results of penguin contributions.
  
 The remainder of this paper is organized as follows. 
 In Sec.~\ref{sec:2}, the theoretical framework is introduced. 
Numerical results and detailed discussions are given in Sec.~\ref{sec:3}. 
Section~\ref{sec:4} has our conclusions.
 

\section{Factorization Amplitudes for Topological Diagrams}\label{sec:2}

\begin{figure} [htb]
\begin{center}
\scalebox{1}{\epsfig{file=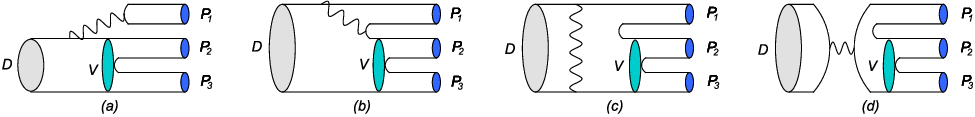}}
\caption{Typical topological diagrams of $D_{(s)} \to P_1 ( V  \to )  P_2 P_3$ under the 
framework of quasi-two-body decay with the wavy line representing a $W$ boson:
  (a) color-favored tree diagram $T$,
  (b) color-suppressed tree diagram $C$, 
  (c) $W$-exchange tree diagram $E$,
  and (d) $W$-annihilation tree diagram $A$.}
\label{tree}
\end{center}
\end{figure}

The quasi-two-body process, $D_{(s)}  \to  P_1 P_2 P_3$ decay is 
divided into two sequential stages. First, the $D_{(s)}$ meson decays into $P_1$
and an intermediate vector resonance $V$. Subsequently,
the vector resonance $V$ decays into the final-state particles $ P_2 P_3$. 
The first decay, $D_{(s)} \to P_1 V$, is a weak decay induced by 
quark-level transition $c \rightarrow \,d(s)\, u \, \bar{d} (\bar s)\, $ 
at leading order in the electroweak interaction, and by 
$c \rightarrow \,u\, \, q\, \bar{q} \, \, (q=u, d, s)$ in penguin diagrams. 
However, since the penguin amplitudes are suppressed by 
Wilson coefficients and Cabibbo-Kobayashi-Maskawa (CKM) matrix elements, 
they can be neglected in the calculation of branching fractions.
The secondary decay $V \to  P_2 P_3$ proceeds via strong interaction 
with the vector resonance described by the RBW model. 
The two-step decays of $D$ meson, $D_{(s)}  \to  P_1 P_2 P_3$, 
can be described by topological diagrams under the framework of 
quasi-two-body decay at tree level, as shown in Fig.~\ref{tree}.
These diagrams include
(a) color-favored emission diagram $T$, 
(b) color-suppressed emission diagram $C$,  
(c) $W$-exchange diagram $E$, and 
(d) $W$-annihilation diagram $A$, which are characterized 
by topological structures of the weak interaction. 
 In these figures, we only illustrate the case that 
the intermediate resonance (labeled by green ovals) is produced 
as a recoiling particle in emission diagrams, 
while the scenario of the intermediate resonance 
being emitted is not shown.

First, we review the intermediate two-body decay amplitudes of 
these four diagrams within the FAT approach. 
Compared to $B$-meson decays, more large nonfactorizable contributions 
are exhibited in $D$ mesons due to the lower energy scale involved.
Even for the $T$ diagram, it is necessary to introduce an additional unknown 
parameter to account for its nonfactorizable effects, as would have been discussed in Ref.~\cite{Zhou:2025}.
A detailed explanation of the parametrization of the amplitudes for the $T, C, E$, and $A$ topologies
can be found in Ref.~\cite{Li:2013xsa} and these amplitudes are expressed as follows:  
\begin{align}\label{2amp}
\begin{aligned}
T_{P_1V}(C_{P_1V})&=\sqrt{2}\, G_{F}\, V_{cq}^*V_{uq^{'}}\, 
a_1(\mu)\, \left( a_2^V(\mu)\right) \, f_V \, m_V\,
F_1^{DP_1}(m_V^2)\, (\varepsilon^{*}_{V} \cdot p_D),
\\
T_{VP_1}(C_{VP_1})&= \sqrt{2}\, G_{F}\, V_{cq}^*V_{uq^{'}}\, 
a_1(\mu)\left(a_2^P(\mu)\right)f_{P_1} m_V
A_0^{DV}(m_{P_1}^2)\, (\varepsilon^{*}_{V} \cdot p_D),
\\
E_{P_1V,VP_1}&=\sqrt{2}\, G_{F}\, V_{cq}^*V_{uq^{'}}\,  
C_2(\mu)\, \chi^E_{q(s)}\, \mathrm{e}^{i\phi^E_{q(s)} }
f_D\, m_V\,  \frac{f_{P_1}\, f_V}{f_\pi \, f_\rho}\, (\varepsilon^{*}_{V} \cdot p_D),
\\
A_{P_1V, VP_1}&=\sqrt{2}\, G_{F}\, V_{cq}^*V_{uq^{'}}\, 
C_1(\mu)\, \chi^A_{q(s)}\, \mathrm{e}^{i\phi^A_{q(s)} }
f_D\, m_V\,  \frac{f_{P_1}\, f_V}{f_\pi \, f_\rho}\, (\varepsilon^{*}_{V} \cdot p_D)\, ,
\end{aligned}
\end{align}
with
\begin{align}\label{WilsonC}
\begin{aligned}
a_1(\mu)&\, =\, C_2(\mu)\, +\, C_1(\mu)\left({1\over N_C}\, +\, \chi^T\right),\\
a_2^{V(P)}(\mu)&\, =\, C_1(\mu)\, +\, C_2(\mu)
\left({1\over N_C}\, +\, \chi_{V(P)}^C \, \mathrm{e}^{i\phi_{V(P)}^C}\right)\, ,
\end{aligned}
\end{align}
where $N_C=3$.
 The scale $\mu$ of the Wilson coefficients is set to the energy release 
 in individual decay modes as indicated by the PQCD approach~\cite{Lu:2000em,Keum:2000wi}:
it is dependent on the mass scales of initial and final states and strong binding energy parameter 
$\Lambda_\mathrm{QCD}$. 
It is defined as
\begin{eqnarray}\label{scalemuTC}
\mu=\sqrt{\Lambda_\mathrm{QCD} m_D(1-r_{P(V)}^2)}\, ,
\end{eqnarray}
for emission diagrams ($T$ and $C$) , and
\begin{eqnarray}\label{scalemuEA}
\mu=\sqrt{\Lambda_\mathrm{QCD} m_D(1-r_P^2)(1-r_V^2)}\, .
\end{eqnarray}
for annihilation diagrams ($E$ and $A$), respectively, 
where $ r_{P(V)}=m_{P(V)}/m_D$ is the mass ratio of the emitted pseudoscalar (vector) meson 
from the weak vertex to the $D$ meson. 

In Eq.~(\ref{2amp}), the subscripts $P_1 V, V P_1$ are used to distinguish between cases 
where the recoiling meson (the first particle) is a pseudoscalar or a vector meson.
The quark in the CKM matrix element is denoted as $q^{(')} =d,s$. 
The polarization vector of vector meson $V$ is represented by $\varepsilon^{*}_{V}$.
The decay constants of the mesons $P_1$ and $V$ are given by $f_{P_1}$ and $f_{V}$, respectively.
The vector form factors for transitions $D_{(s)} \to P_1$ and $D_{(s)} \to V$ 
are denoted by $F_{1}^{DP_1}$ and $A_{0}^{DV}$, respectively.
The parameters $\chi^T,\, \chi_{V(P)}^C , \, \chi_{q(s)}^E$ , and $ \chi_{q(s)}^A $ 
are related to the magnitudes of topological diagrams T, C, E and A, respectively,
associated with their corresponding strong phases $ \phi_{V(P)}^C , \, \phi_{q(s)}^E$, and $\phi_{q(s)}^A $,
needed to be fitted globally from the experimental data.
In order to minimize the number of free parameters, we do not introduce 
a strong phase for the $T$ diagram, 
as we only consider its phase relative to other diagrams.
For the C diagram, we introduce two distinct sets of parameters: $ \chi_{V}^C,  \phi_{V}^C$ 
and $ \chi_{P}^C,  \phi_{P}^C$, to account for whether the emitted meson 
is a vector ($V$) or a pseudoscalar ($P$) particle.
Additionally, the subscripts $q$ and $s$ in $ \chi_{q(s)}^E, \, \chi_{q(s)}^A ,\phi_{q(s)}^E, \, $ and $\phi_{q(s)}^A$
are used to differentiate between the strongly produced light-quark pair (u or d) and the strange-quark pair.
This distinction, which indicates the $SU(3)$ breaking effect, is  crucial for accurately describing the 
dynamics of the decay processes of $E$ and $A$ diagrams. 
For the two annihilation amplitudes, it is also noted that 
the Glauber phase $S_\pi $ is introduced  
as an additional strong phase arising from their nonfactorizable contributions,
when a pion is involved in the final state~\cite{Li:2012cfa, Li:2013xsa}.
This strong phase plays a more crucial role in the interference between the emission diagrams 
and the annihilation ones, particularly with the availability of more precise experimental data~\cite{Zhou:2025}.

The secondary subprocess $V \to P_2 \, P_3$ is described using the RBW line shape, 
which is widely applied in both experimental data analyses and theoretical modeling of resonances, 
$\rho$,\, $K^*$,\, $\omega$, and $\phi$ in the strong decay 
process~\cite{BESIII:2021xox,BESIII:2022ewq,BESIII:2022vaf,
Li:2018psm,Ma:2019qlm,Fan:2020gvr,Ma:2020jsb,Wang:2020nel,Wang:2024enc}.
The explicit expression of RBW distribution is written as
\begin{align}\label{RBW}
L^{\mathrm{RBW}}(s)=\frac{1}{s-m_{V}^{2}+i m_{V} \Gamma_{V}(s)}\, ,
\end{align}
where $s=m^2_{23}=(p_2+p_3)^2$ is the invariant mass squared of the two-body system, with
$p_2$ and $p_3$ denoting the 4-momenta of the two mesons $P_2$ and $P_3$ moving collinearly, 
respectively. 
The energy-dependent width of the vector resonance $\Gamma_{V}(s)$ is defined as
\begin{align}\label{width}
\Gamma_{V}(s)=\Gamma_{0}\left(\frac{q}{q_{0}}\right)^{3}\left(\frac{m_{V}}{\sqrt{s}}\right) 
X^{2}\left(q\, r_{\mathrm{BW}}\right)\, ,
\end{align}
where $q$ is the momentum magnitude of either $P_2$ or $P_3$ 
in the rest frame of the resonance $V$, given by
\begin{align}
q=\frac{1}{2} \sqrt{\left[s-\left(m_{P_2}+m_{P_3}\right)^{2}\right]
 \left[s-\left(m_{P_2}-m_{P_3}\right)^{2}\right] / s}\, ,
 \end{align}
 and $q_0$ is the value of $q$ when $s = m^2_{V}$.
The Blatt-Weisskopf barrier factor $X\left(q\, r_{\mathrm{BW}}\right)$ is defined as~\cite{Blatt:1952ije}
\begin{equation}
X\left(q\, r_{\mathrm{BW}}\right)
=\sqrt{[1+\left(q_{0}\, r_{\mathrm{BW}}\right)^{2}]/[1+\left(q\, r_{\mathrm{BW}}\right)^{2}}]\, .
\end{equation}
In cases where the pole mass $m_V$ lies outside the kinematics region, 
i.e. $m_{V}<m_{P_2}+m_{P_3}$, the mass $m_V$ is replaced with 
an effective mass $m^{\mathrm{eff}}_{V}$, calculated using the $ad$ $hoc$ 
formula~\cite{Aaij:2014baa, Aaij:2016fma}
\begin{align}\label{massDstar}
 m_{V}^{\text {eff }}\left(m_{V}\right)=m^{\min }+\left(m^{\max }-m^{\min }\right)
 \left[1+\tanh \left(\frac{m_{V}-\frac{m^{\min }+m^{\max }}{2}}{m^{\max }-m^{\min }}\right)\right]\, ,
 \end{align}
 where $m^{\max }$ and $m^{\min }$ are the upper and lower boundaries of the kinematic region, respectively.
 The barrier radius $r_{\mathrm{BW}}=4.0 (\mathrm{GeV})^{-1}$ is used for all resonances~\cite{LHCb:2019sus}.
 The masses $m_V$ and full widths $\Gamma_{0}$ of the resonances
 are taken from Particle Data Group (PDG)~\cite{ParticleDataGroup:2024cfk}
 and are listed in Table~\ref{tab:mass and width}.
\begin{table}[tbhp]
\caption{Masses $m_V$ and full widths $\Gamma_0$   
of vector resonant states. }
\vspace{3mm}
\label{tab:mass and width}
\centering
\begin{tabular}{cccc}
\hline
Resonance ~~&~~ Line shape Parameters (MeV)~~ &~~Resonance ~~&~~ Line shape Parameters (MeV)
\\ \hline
$\rho(770)~~$  & $m_V\, =\, 775.26\,  \, $ &
$\omega(782)~~$  & $m_V\, =\, 782.65\,  \, $ \\
& $\Gamma_0\, =\, 149.1\, \, $
& &$\Gamma_0\, =\, 8.49\, \, $\\
$K^*(892)^+~~$  & $m_V\, =\, 891.66\,  \, $ 
&$K^*(892)^0~~$  & $m_V\, =\, 895.55\,  \, $\\
& $\Gamma_0\, =\, 50.8\, \,  $
 & &$\Gamma_0\, =\, 47.3\, \, $\\
$\phi(1020)~~$ & $m_V\, =\, 1019.46\,  \, $ \\
&  $\Gamma_0\, =\, 4.25 \, \, $\\
\hline
\end{tabular}
\end{table}

Finally the intermediate resonance $V$ described above decays via strong interaction,
with the matrix element $\left\langle  P_2 \left(p_{2}\right) P_3 \left(p_{3}\right) | V (p_V)\right\rangle$
parametrized as a strong coupling constant $g_{ V P_2 P_3}$.
This coupling constant can be extracted from the measured 
partial decay widths $\Gamma_{V  \to P_2 P_3}$ using the relations
\begin{align}
\Gamma_{V \rightarrow P_{2} P_{3}}
=\frac{2}{3} \frac{p_{c}^{3}}{4 \pi m_{V}^{2}} g_{V \rightarrow P_{2} P_{3}}^{2}\, ,
\end{align}
where $p_c$ is the magnitude of pseudoscalar meson momentum in the rest frame of the vector meson. 
The numerical values of the strong coupling constants $g_{\rho \rightarrow \pi^{+} \pi^{-}}$,
 $g_{K^{*} \rightarrow K^{+} \pi^{-}}$ and $g_{\phi \rightarrow K^{+} K^{-}}$ have already been 
 determined  from experimental data~\cite{Cheng:2013dua}
\begin{align}\label{gVPP}
&g_{\rho \rightarrow \pi^{+} \pi^{-}}=6.0\, ,
\quad g_{K^{*} \rightarrow K^{+} \pi^{-}}=4.59\, ,
\quad g_{\phi \rightarrow K^{+} K^{-}}=-4.54\, .
\end{align}
Other strong coupling constants can be derived from these results 
using relationships based on the quark model result~\cite{Bruch:2004py}.
These relationships are
$$g_{\rho \rightarrow K^{+} K^{-}}: g_{\omega \rightarrow K^{+} K^{-}}: g_{\phi \rightarrow K^{+} K^{-}}=1: 1:- \sqrt{2},$$
$$g_{\rho^{0} \pi^{+} \pi^{-}} = g_{\rho^+ \pi^0 \pi^+}\, , \, g_{\rho^{0} \pi^{0} \pi^{0}}= g_{\omega \pi^+ \pi^-} =0  \, ,$$
$$g_{\rho^{0} K^{+} K^{-}}=-g_{\rho^{0} K^{0} \bar{K}^{0}}=g_{\omega K^{+} K^{-}}=g_{\omega K^{0} \bar{K}^{0}}, \, 
g_{\phi K^{+} K^{-}}=g_{\phi K^{0} \bar{K}^{0}}\, .$$

Combining the two subprocesses together for $D_{(s)}  \to  P_1 (V \to ) P_2 P_3$,  
the decay amplitudes of the four topological diagrams shown in Fig.~\ref{tree} are listed
 in the following: 
\begin{align}\label{eq:T}
\begin{aligned}
T_{(P_2 P_3)P_1}
&=\left\langle P_2 \left(p_{2}\right) P_3 \left(p_{3}\right) \left|(\bar{q} c)_{V-A}\right| D (p_D) \right\rangle
\left \langle P_1 \left(p_{1}\right)\left|(\bar{u} q^{\prime})_{V-A}\right| 0\right\rangle\\
&=\frac{\left\langle  P_2 \left(p_{2}\right) P_3 \left(p_{3}\right) | V (p_V)\right\rangle}{s-m_{V}^{2}+i m_{V} \Gamma_{V}(s)}
\left\langle V (p_V) \left|(\bar{q} c)_{V-A}\right| D(p_D) \right\rangle\left \langle P_1 \left(p_{1}\right)\left|(\bar{u} q^{\prime})_{V-A}\right| 0\right\rangle\\
&=2\, \mathbf{p_1} \cdot \mathbf{p_2}\, \sqrt 2 G_{F} V_{cq}^*V_{uq^{'}}\, \, a_1({\mu})\, f_{P_1}\, m_{V} A_0^{D V} (m_{P_1}^2)\, \frac{g_{ V P_2 P_3}} {s-m_{V}^{2}+i m_{V} \Gamma_{V}(s)} \, ,\\
T_{P_1(P_2 P_3)}&=\left\langle P_2 \left(p_{2}\right) P_3 \left(p_{3}\right)  |(\bar{u} q^{\prime})_{V-A} |0 \right \rangle
\left\langle P_1 (p_1)|(\bar q c)_{V-A}| D(p_D) \right \rangle \\
&=\frac{\left\langle  P_2 \left(p_{2}\right) P_3 \left(p_{3}\right) | V (p_V)\right\rangle}{s-m_{V}^{2}+i m_{V} \Gamma_{V}(s)}
\left\langle V (p_V) \left|(\bar{q} c)_{V-A}\right| 0 \right\rangle\left \langle P_1 \left(p_{1}\right)\left|(\bar{u} q^{\prime})_{V-A}\right| D(p_D)\right\rangle\\
&=2 \, \mathbf{p_1} \cdot \mathbf{p_2}\,  \sqrt 2 G_{F} V_{cq}^*V_{uq^{'}}\,  \, a_1({\mu})\, f_{V} m_{V} F_1^{D P_1} (s)\,  \frac{g_{ V P_2 P_3}} {s-m_{V}^{2}+i m_{V} \Gamma_{V}(s)} \, ,
\end{aligned}
\end{align}
\begin{align}\label{eq:C}
\begin{aligned}
C_{(P_2 P_3) P_1}&=\left\langle P_2 \left(p_{2}\right) P_3 \left(p_{3}\right) 
\left|(\bar{q}^{\prime} c)_{V-A}\right| D (p_D) \right\rangle
\left \langle P_1 \left(p_{1}\right)\left|(\bar{u} q)_{V-A}\right| 0\right\rangle\\
&=\frac{\left\langle  P_2 \left(p_{2}\right) P_3 \left(p_{3}\right) | V (p_V)\right\rangle}{s-m_{V}^{2}+i m_{V} \Gamma_{V}(s)}
\left\langle V (p_V) \left|\bar{q}^{\prime} c)_{V-A}\right| D(p_D) \right\rangle\left \langle P_1 \left(p_{1}\right)
\left|(\bar{u} q)_{V-A}\right| 0\right\rangle\\
&=2\, \mathbf{p_1} \cdot \mathbf{p_2}\, \sqrt 2 G_{F} V_{cq}^*V_{uq^{'}}\,  
a_2^P(\mu)\,
f_{P_1}\, m_{V} A_0^{DV} (m_{P_1}^2)\, \frac{g_{ V P_2 P_3}} {s-m_{V}^{2}+i m_{V} \Gamma_{V}(s)} \, ,\\
C_{P_1 (P_2 P_3) }&=\left\langle P_2 \left(p_{2}\right) P_3 \left(p_{3}\right) |(\bar{u} q)_{V-A} |0 \right \rangle
\left\langle P_1 (p_1)|(\bar{q}^{\prime} c)_{V-A}| D(p_D) \right \rangle \\
&=\frac{\left\langle  P_2 \left(p_{2}\right) P_3 \left(p_{3}\right)| V (p_V)\right\rangle}{s-m_{V}^{2}+i m_{V} \Gamma_{V}(s)}
\left\langle V (p_V) \left|(\bar{u} q)_{V-A}\right| 0 \right\rangle\left \langle P_1 \left(p_{1}\right)\left|(\bar{q}^{\prime} c)_{V-A}\right| D(p_D)\right\rangle\\
&=2\, \mathbf{p_1} \cdot \mathbf{p_2}\,  \sqrt 2 G_{F} V_{cq}^*V_{uq^{'}}\,  a_2^V(\mu)\, 
  f_{V} m_{V} F_1^{D P_1} (s)\,\frac{g_{ V  P_2 P_3}} {s-m_{V}^{2}+i m_{V} \Gamma_{V}(s)} \, ,
\end{aligned}
\end{align}
\begin{align}\label{eq:E}
\begin{aligned}
E_{P_1 (P_2 P_3)}&=\left\langle P_1 \left(p_{1}\right) P_2 \left(p_{2}\right) P_3 (p_3) \left|\mathcal{H}_{eff}\right| D(p_D) \right\rangle\\
&=\frac{\left\langle  P_2 \left(p_{2}\right) P_3 \left(p_{3}\right) | V (p_V)\right\rangle}{s-m_{V}^{2}+i m_{V} \Gamma_{V}(s)}
\left\langle V (p_V) P_1(p_1) \left|\mathcal{H}_{eff} \right| D(p_D)\right\rangle\\
&=2\, \mathbf{p_1} \cdot \mathbf{p_2}\,  \sqrt 2 G_{F} V_{cq}^*V_{uq^{'}}\, C_2(\mu)\, 
\chi^E_{q(s)}\, \mathrm{e}^{i\phi^E_{q(s)} } \, 
f_{D} m_{V} \frac{f_{V} f_{P_1}}{f_{\pi} \, f_\rho}\, \frac{g_{V  P_2 P_3}} {s-m_{V}^{2}+i m_{V} \Gamma_{V}(s)} \, ,\\
A_{P_1 (P_2 P_3)}&=\left\langle P_1 \left(p_{1}\right) P_2 \left(p_{2}\right) P_3 (p_3) \left|\mathcal{H}_{eff}\right| D(p_D) \right\rangle\\
&=\frac{\left\langle  P_2 \left(p_{2}\right) P_3 \left(p_{3}\right) | V (p_V)\right\rangle}{s-m_{V}^{2}+i m_{V} \Gamma_{V}(s)}
\left\langle V (p_V) P_1(p_1) \left|\mathcal{H}_{eff} \right| D(p_D)\right\rangle\\
&=2\, \mathbf{p_1} \cdot \mathbf{p_2}\,  \sqrt 2 G_{F} V_{cq}^*V_{uq^{'}}\, 
C_1(\mu)\, \chi^A_{q(s)}\, \mathrm{e}^{i\phi^A_{q(s)} }
f_{D} m_{V} \frac{f_{V} f_{P_1}}{f_{\pi} \, f_\rho}\, \frac{g_{V  P_2 P_3}} {s-m_{V}^{2}+i m_{V} \Gamma_{V}(s)} \, ,
\end{aligned}
\end{align}
where the prefactor $2\, \mathbf{p_1} \cdot \mathbf{p_2}$ in these amplitudes is derived 
in the rest frame of the $P_2 P_3$ system.
In the decay process $D_{(s)}  \to  P_1 (V \to ) P_2 P_3$, 
where $p_V=p_2+p_3=\sqrt{s}$, the form factor $F_{1}^{D P_1}(s)$ depends on
the invariant mass $s$ of the $P_2 P_3$ system. This is in contrast to two-body decays,
where the form factor is evaluated at a fixed value of $s$ 
(typically $s=m_V^2$, the mass of the vector meson).

The matrix element for the decay $D_{(s)}  \to  P_1 P_2 P_3$ can be expressed 
in the form:
\begin{align}\label{Ampmatrix element}
\left\langle P_1 (p_{1})  P_2 (p_2) P_3 (p_{3}) \left| \mathcal{H}_{eff} \right| D_{(s)}(p_D) \right\rangle
= 2\, \mathbf{p_1} \cdot \mathbf{p_2}\,  \mathcal{A}(s)\, ,
\end{align}
where $\mathcal{A}(s)$ represents the summation of amplitudes in Eqs.~(\ref{eq:T})-(\ref{eq:E}) 
with the prefactor $2\, \mathbf{p_1} \cdot \mathbf{p_2}$ factored out.
The differential decay width for $D_{(s)}  \to  P_1 P_2 P_3$ is given by
\begin{align}\label{dwidth}
\begin{aligned}
d \, \Gamma=&d\, s  \,  \frac{1}{(2 \pi)^3}\,
\frac{(\left|\mathbf{p_1} \| \mathbf{p}_{2}\right|\,)^3}{6 m_D^3}\, | \mathcal{A}(s)|^2\, , \\
\end{aligned}
\end{align}
where the magnitudes $|\mathbf{ p_{1}} |$ and $|\mathbf{ p_{2}} |$ of the
three-momenta of the particles $P_1$ and $P_2$ are evaluated in
the rest frame of the $P_2 P_3$ system. 
The expression for $|\mathbf{ p_{1}} |$ is
\begin{align}
\begin{aligned}
|\mathbf{ p_{1}} |=&\frac{1}{2\, \sqrt s}\,
\sqrt{ \left[(m_D^2-m_{P_1}^2)^2 \right] \, -2(m_D^2+m_{P_1}^2) \,s +s^2}\, ,\\
\end{aligned}
\end{align}
and $|\mathbf{ p_{2}} |=q$, corresponding to the momentum of $P_2$ (or $P_3$),
in the $P_2 P_3$ system rest frame, as defined in Eq.~(\ref{width}).
\section{Numerical Results of Branching Fractions}\label{sec:3}
The input parameters are categorized as follows: 
\begin{itemize}
 \item \textbf{Electroweak coefficients:} 
These include the CKM
matrix elements and Wilson coefficients.
\begin{itemize}
 \item The CKM matrix elements are provided in the PDG~\cite{ParticleDataGroup:2024cfk}.
 \item The Wilson coefficients $C_{1(2)}(\mu)$ [see Eq.~(\ref{WilsonC})] for $D$ meson decays 
 are taken from Eqs.~(B1), (B2), and (B7) in the Appendix of Ref.~\cite{Li:2012cfa}.
 The scale $\mu$ in $C_{1(2)}(\mu)$ is dependent on masses involved in each decay mode
 and the soft scale in the $D$ meson, $\Lambda_\mathrm{QCD}$, as 
 in Eqs.~(\ref{scalemuTC}) and (\ref{scalemuEA}).
\end{itemize}
 \item \textbf{Hadronic parameters:} The parameters $m_V$, $\Gamma_0$ 
involved in the strong interaction decays of vector mesons have been 
listed in Table~\ref {tab:mass and width}. 
The strong coupling constants $g_{V P_2 P_3}$ are presented in Eq.~(\ref{gVPP}),
along with the associated relationships provided below.
 \item \textbf{Nonperturbative QCD parameters:} These encompass decay constants, transition form factors 
and nonfactorizable parameters $\chi^T, \chi_{V(P)}^C, \chi^E_{q(s)}, \chi^A_{q(s)}$ and $\phi_{V(P)}^C,
\phi^E_{q(s)}, \phi^A_{q(s)}, S_\pi$.
\begin{itemize}
\item The decay constants of light pseudoscalar mesons and vector mesons 
 (in units of MeV) are listed in Table~\ref{tab:decay constants}. 
 The decay constants of $\pi, K$, and $D_{(s)}$
 are obtained from the PDG~\cite{ParticleDataGroup:2024cfk}. 
 For the decay constants of vector mesons that have not been experimentally measured, 
 we adopt the same theoretical values as used in quasi-two-body decays of 
 $B$ meson~\cite{Zhou:2023lbc,Zhou:2024qmm}, keeping a 5\% error.
\begin{table}[tbhp]
\caption{ Decay constants of pseudoscalar and vector mesons. }
\vspace{3mm}
\label{tab:decay constants}
\newsavebox{\tablebox}
\begin{lrbox}{\tablebox}
\centering
\begin{tabular}{ccccccccccc|}
\hline
$f_{\pi}$ & $f_{K}$  &  $f_{D}$ & $f_{D_s}$ &$f_{\rho}$ &$f_{K^*}$ & $f_{\omega}$&  $f_{\phi}$ &
\\ \hline
$130.2 \pm 1.7$ & $155.6 \pm 0.4$ & $211.9 \pm 1.1$ & $258 \pm 12.5$ & $213 \pm 11$ & $220 \pm 11$ & $192\pm 10 $ & $225\pm11$ &
\\
\hline
\end{tabular}
\end{lrbox}
 \scalebox{0.9}{\usebox{\tablebox}}
\end{table}
\item The transition form factors of $D$-meson decays, specifically $D \to \pi$ and $D \to K$,
have been measured by several experiments, including CLEO-c~\cite{CLEO:2009svp},
Belle~\cite{Belle:2006idb}, and BESIII~\cite{BESIII:2018xre, BESIII:2024zft}.
While most form factors have not been measured in experiments, 
theoretical predictions are available from various theoretical methods, 
such as lattice QCD~\cite{Bernard:2009ke}, QCD sum rule~\cite{Du:2003ja,Tian:2024lrn},
quark model~\cite{Melikhov:2000yu}, covariant light-front quark model~\cite{Chen:2009qk,Verma:2011yw,Wang:2008ci},
improved light-cone harmonic oscillator model~\cite{Hu:2024tmc},
heavy meson and chiral symmetries ($\mathrm{HM_\chi T}$)~\cite{Fajfer:2005ug},
 hard-Wall AdS/QCD model~\cite{Momeni:2022gqb},
and four-flavor holographic QCD~\cite{Ahmed:2023pod}.
 Since different theoretical approaches yield varying values of the form factors,
 we adopt their values at zero recoil
momentum ($Q^2=0$) within the range of experimental and theoretical results, 
and assign a 10\% error bar to account for uncertainties. 
As will be demonstrated in the following section, 
the form factor uncertainty constitutes the major source of theoretical uncertainty 
in our calculation.
 These values, along with $\alpha_i$ parameters, are presented in Table~\ref{tab:ff}.
 The $Q^2$ dependence of these form factors is uniformly
parametrized using a dipole form, as follows:
 \begin{equation}\label{eq:ffdipole}
F_{i}(Q^{2})
={F_{i}(0)\over 1-\alpha_{1}{Q^{2}\over m_{\rm pole}^{2}}+\alpha_{2}{Q^{4}\over m_{\rm pole}^{4}}},
\end{equation}
where $F_{i}$ represents form factor $F_{1}$ or $A_{0}$, and $m_{\rm pole}$ is the mass of the 
corresponding pole state, e.g. $m_{D^*}$ for $F_{1}^{D \pi, D \eta^{(')}, D_s K}$, 
$m_{D_s^{*}}$ for $F_{1}^{DK,D_s \eta^{(')}}$, $m_D$ for $A_{0}^{D \rho, D \omega, D_s K^*}$,
and $m_{D_s}$ for $A_{0}^{D K^*, D_s \phi}$. 
The dipole model parameters $\alpha_{1,2}$ can be found in Table II of~\cite{Fu-Sheng:2011fji}.
\begin{table} [hbt]
\caption{Form factors and dipole model parameters. }\label{tab:ff}
\vspace{3mm}
\centering
\begin{tabular}{|c|c|c|c|c|c|c|c|c|c|c|}
\hline&
$~~F_{1}^{D\to\pi}~~$&
$~~F_{1}^{D\to K}~~$&
$~~F_{1}^{D_s\to K}~~$&
$~~F_{1}^{D\to \eta_q}~~$&
$~~F_{1}^{D\to \eta_s}~~$\\
\hline
$~F_i(0)~$&
0.60&
0.74&
0.66 &
0.76&
0.79\\
$\alpha_1$&
1.24&
1.33&
1.20&
1.03&
1.23\\
$\alpha_2$&
0.24&
0.33&
0.20&
0.29&
0.23\\
\hline&
$~~A_{0}^{D\to \rho}~~$&
$~~A_{0}^{D\to K^*}~~$ &
$~~A_{0}^{D_s\to K^*}~~$ &
$~~A_{0}^{D\to \omega}~~$&
$~~A_{0}^{D_s \to \phi}~~$ \\
\hline
$~F_i(0)~$&
0.84&
0.73&
0.68&
0.68&
 0.70  \\
$\alpha_1$&
1.36&
1.17&
1.20&
1.36&
 1.10 \\
$\alpha_2$&
0.36&
0.17&
0.20&
0.36&
0.10 \\
\hline
\end{tabular}
\end{table}
\item The nonfactorizable parameters of the topological diagram amplitudes 
in Eqs.~(\ref{eq:T})-(\ref{eq:E}) consist of 15 free parameters:
$\chi^T,\, \chi_{V(P)}^C , \, \chi_{q(s)}^E, \text{and}\, \chi_{q(s)}^A $, along with
their associated phases $\phi_{V(P)}^C , \, \phi_{q(s)}^E, \, \phi_{q(s)}^A$,  
$S_\pi$ [the factor $\mathrm{e}^{i\, S_\pi}$ multiplies the 
$E(A)_{P_1(P_2P_3)}$ terms in Eq.~(\ref{eq:E}) ],  
and the soft scale $\Lambda_{\mathrm{QCD}}$ [see Eqs.~(\ref{scalemuTC}) and (\ref{scalemuEA})]. 
These parameters
are extracted through a global fit to 41 experimental data of $D \to PV$  with significance exceeding $3 \sigma$, 
as updated in Ref.~\cite{Zhou:2025}. The best-fitted parameters together with their corresponding uncertainties are
\begin{align}\label{parameter}
\begin{aligned}
\chi^{T}=-0.29 \pm 0.01,~~~& \\
\chi_P^{C}=-0.47 \pm 0.02,~~~&\phi_P^{C}=0.37 \pm 0.12,\\
\chi_V^{C}=-0.41 \pm 0.01,~~~&\phi_V^{C}=-0.52\pm 0.02, \\
\chi_q^{E}=0.13\pm0.01,~~~&\phi_q^{E}=2.68\pm 0.10, \\
\chi_s^{E}=0.24\pm0.01,~~~&\phi_s^{E}=3.52\pm 0.07, \\
\chi_q^{A}=0.10\pm0.003,~~~&\phi_q^{A}=-2.08\pm 0.16, \\
\chi_s^{A}=0.18\pm0.01,~~~&\phi_s^{A}=2.62\pm 0.10, \\
\Lambda_\mathrm{QCD}=(0.24\pm0.01)~&\mathrm{GeV},S_\pi=-1.80\pm 0.17,
\end{aligned}
\end{align}
with the fitted $\chi^2/\mathrm{d.o.f.}=6.2$. 
These nonfactorizable parameters are highly precise, with the exception 
of the strong phase $\phi_P^{C}$.
As is known, $\phi_P^{C}$ is primarily constrained by data from modes
dominated by the $C_{VP}$ diagram.
Specifically, those involving only the $C_{VP}$ term or 
accompanied by minor contributions from annihilation diagrams.
However, the experimental data for such modes remain scarce, 
limiting the precision of $\phi_P^{C}$.
Using the fitted parameters $\chi^{T}\, , \chi_{V(P)}^{C}\, , \phi^{C}_{V(P)}$, 
we calculate the values of $a_1(\mu)$ and $a_2^{V(P)}(\mu)$ 
for the emission diagrams of each decay modes.
For instance, in the decay $D^+ \to \bar K^0 \, (\rho^+) \pi^+ \pi^-$,
we determine $a_1(\mu)\, , a_2^{V}(\mu)$,
alongside the corresponding Wilson coefficients of $C_{1(2)}(\mu)$ 
at fixed scale $\mu$ relevant to this process,
\begin{align}\label{parameter}
\begin{aligned}
\mu=0.61\, \mathrm{GeV},& ~ \\
C_1(\mu)=-0.81,~~~&C_2(\mu)=1.49,\\
a_1(\mu)=-1.45,~~~&a_2^V(\mu)=1.09 \, \mathrm{e}^{i\, 0.26}.
\end{aligned}
\end{align}
 \end{itemize}
\end{itemize}

The branching fractions of $D_{(s)}  \to  P_1 (V \to ) P_2 P_3$ 
can be obtained by integrating the differential width given in Eq.~(\ref{dwidth}) 
over the region of $s$. 
Our numerical results for the branching fractions of $D_{(s)}$ decays 
are collected in Tables~\ref{Prho}-\ref{Pomega}, corresponding to the processes
$D_{(s)} \to P_1 \rho \to P_1 \pi \pi$, $D_{(s)} \to P_1 K^{*} \to P_1 K \pi$, 
$D_{(s)} \to P_1 \phi \to P_1 K\bar K$, and 
$D_{(s)} \to P_1 \omega \to P_1 K\bar K$, respectively. 
Each table is further categorized into three decay modes: 
Cabibbo-favored (CF), singly Cabibbo-suppressed (SCS), and doubly Cabibbo-suppressed (DCS) modes.
The tables additionally specify the intermediate resonance decays 
and topological contributions, with symbols such as $T_{PV,PV},C_{PV,PV}, E, A$
 denoting these contributions
to facilitate the analysis of branching fraction hierarchies.
 In the subsequent discussions, we analyze each table in turn. 
 For our results in the FAT approach, the first uncertainty arises from the nonperturbative parameters 
in Eq.~(\ref{parameter}), while the second and third uncertainties are estimated 
by varying the form factors by 10\% and unmeasured decay constants by  5\%,
respectively. Notably, the dominant source of 
uncertainty stems from the form factors. 
In the last column, we list the experimental data for comparison, 
while for unmeasured cases, the predictions of branching fractions 
can be tested against future experimental data.

\subsection{Branching fractions of $D_{(s)} \to P_1\, ( \rho \to)\,  \pi \pi$ and  $D_{(s)} \to P_1\, (K^{*} \to) \, K \pi$}

\begin{table}[!hb]
\caption{Branching Fractions in the FAT approach of quasi-two-body decays 
$D_{(s)} \to P_1 (\rho \to)  \pi \pi$ ($\times 10^{-3}$), 
(with the uncertainties from the fitted parameters, 
form factors and decay constants, respectively,) together with
the experimental data~\cite{ParticleDataGroup:2024cfk}.
In the second column, the characters $T_{PV,PV},C_{PV,PV}, E, A$ represent
the corresponding topological diagram contributions.}
 \label{Prho}
\begin{center}
\begin{tabular}{cccc}
 \hline \hline
{Modes}     &  Amplitudes & FAT results       &  Experimental data   \\
\hline
CF & $V_{cs}^*V_{ud}\, $ & &\\
$D^0 \to K^-(\rho^+\to)\pi^+ \pi^0$ & $T_{PV}, E$ & $89.78 \pm4.14\pm21.12\pm9.08$  &$ 112\pm7 $\\
                           
$D^0 \to \bar K^0(\rho^0\to)\pi^+ \pi^-$ & $C_{VP}, E$ &$9.30 \pm1.17\pm2.30\pm0.32$  &$ 12.6\pm 1.6 $ \\ 

$D^+ \to \bar K^0 (\rho^+\to)\pi^+ \pi^0$ & $T_{PV}, C_{VP} $ &$90.61 \pm9.51\pm36.00\pm16.42$  &$122.8\pm 12  $\\

$D^+ \to \eta (\rho^+\to)\pi^+ \pi^0$ & $T_{PV}, C_{VP}, A $ & $0.18 \pm0.14\pm0.13\pm0.04$  &$  $\\
   
 $D^+ \to \eta^\prime (\rho^+\to)\pi^+ \pi^0$ & $T_{PV}, C_{VP}, A $ &$1.68 \pm0.05\pm0.23\pm0.10$  &$  $\\
                                  
 $D_s^+ \to \eta (\rho^+\to)\pi^+ \pi^0$ & $T_{PV}, A $ &$79.21 \pm3.34\pm16.63\pm7.92$  &$89\pm8  $\\
   
 $D_s^+ \to \eta^\prime (\rho^+\to)\pi^+ \pi^0$ & $T_{PV}, A $ &$34.66 \pm0.98\pm6.59\pm3.46$  &$ 58\pm15 $\\
 
 \hline 
  SCS  &$V_{c d(s)}^*V_{ud(s)}\, $ &&\\
 $D^0 \to \pi^+ (\rho^-\to)\pi^- \pi^0$ & $T_{VP}, E$ &$4.37\pm0.30\pm 0.72\pm0.11 $   &$5.15 \pm 0.25$\\
  
$D^0 \to \pi^-(\rho^+\to)\pi^+ \pi^0$ & $T_{PV}, E$  &$8.55 \pm0.44\pm1.50\pm0.86$  &$ 10.1\pm0.4 $\\

$D^0 \to \pi^0(\rho^0\to)\pi^+ \pi^-$ & $C_{PV,VP}, E$ &$2.65 \pm0.24\pm0.30\pm0.18$  &$ 3.86\pm0.23 $ \\  

$D^0 \to \eta (\rho^0\to)\pi^+ \pi^-$ & $C_{PV,VP}, E $ &$0.42\pm0.10\pm0.08\pm0.00$  &$ $\\

$D^0 \to \eta^{\prime} (\rho^0\to)\pi^+ \pi^-$ & $C_{PV,VP}, E $ & $0.24\pm0.01\pm0.03\pm0.01$  &$ ~~$\\           
                             
$D^+ \to \pi^+ (\rho^0\to)\pi^+ \pi^-$ & $T_{VP}, C_{PV},A$ &$0.48 \pm0.04\pm0.14\pm0.02$ & $ 0.83\pm 0.14$\\    
  
$D^+ \to \pi^0 (\rho^+\to)\pi^+ \pi^0$ & $T_{PV}, C_{VP},A$ &$2.90 \pm0.24\pm1.05\pm0.49$  &$ $\\                                             
                        
$D_s^+ \to K^+ (\rho^0\to)\pi^+ \pi^-$ & $C_{PV}, A $ & $1.59 \pm0.10\pm0.32\pm0.16$  &$2.17 \pm 0.25 $\\    
  
$D_s^+ \to K^0 (\rho^+\to)\pi^+ \pi^0$ & $T_{PV}, A $ &$8.16 \pm0.29\pm1.70\pm0.82$  &$ 5.46\pm 0.95 $\\   
                                 
 \hline
  DCS &$V_{cd}^*V_{us}\,$ &&\\
 $D^0 \to K^+ (\rho^-\to)\pi^- \pi^0$ & $T_{VP}, E$ & $0.12 \pm0.01\pm0.03\pm0.00 $   &$  $\\
  
$D^0 \to K^0(\rho^0\to)\pi^+ \pi^-$ & $C_{VP}, E $ & $0.03 \pm0.00\pm0.01\pm0.00$  &$  $ \\  
  
 $D^+ \to K^+ (\rho^0\to)\pi^+ \pi^-$ & $T_{VP}, A $  &$0.23 \pm0.01\pm0.05\pm0.00$  &$ 0.19 \pm 0.05$\\    
  
$D^+ \to K^0 (\rho^+\to)\pi^+ \pi^0$ & $C_{VP}, A $  &$0.19 \pm0.02\pm0.04\pm0.00$  &$ $\\   
                       
 \hline \hline
\end{tabular}
\end{center}
\end{table}

\begin{table}[tbhp]
\caption{Same as Table~\ref{Prho} for quasi-two-body decays 
$D_{(s)} \to P_1 (K^{*} \to ) K \pi$  in units of $10^{-3}$.}
 \label{PKstar}
\begin{center}
\begin{tabular}{cccc}
 \hline \hline
Decay Modes     &  Amplitudes & FAT  results       &  Experimental data   \\
\hline
CF & $V_{cs}^*V_{ud}\, $ & &\\

$D^0 \to \pi^+ ({K}^{*-}\to)K^- \pi^0 $ &$T_{VP}, E$ & $18.84\pm1.43\pm3.00\pm0.52$ &$ 19.5 \pm 2.4$ \\
 
 $D^0 \to \pi^0 (\bar{K}^{*0}\to)K_S^0 \pi^0 $ &$C_{PV}, E$ & $5.67\pm0.50\pm0.98\pm0.57$ &$ 8.1\pm0.7$   \\  
 
 $D^0 \to \eta (\bar{K}^{*0}\to)K_S^0 \pi^0 $ &$C_{PV}, E$ & $3.01\pm0.34\pm0.60\pm0.30$ &$2.9\pm0.7 $   \\ 
 
 $D^+ \to \pi^+ (\bar{K}^{*0}\to) K^- \pi^+ $ &$T_{VP}, C_{PV}$ & $11.17\pm0.92\pm3.49\pm1.72$ &$ 10.4\pm1.2$   \\    

 $D_s^+ \to K^+ (\bar{K}^{*0}\to)K^- \pi^+ $ &$C_{PV}, A$ & $25.00\pm1.64\pm6.73\pm2.65$ &$25.8\pm0.6 $   \\   

 $D_s^+ \to \bar K^0 ({K}^{*+}\to) K^+ \pi^0 $ &$C_{VP}, A$ & $5.22\pm0.73\pm1.26\pm0.16$ &$ $   \\  
 
 \hline
   SCS &$V_{c d(s)}^*V_{ud(s)}\, $ &&\\

  $D^0 \to K^+ ({K}^{*-}\to)K^- \pi^0 $ &$T_{VP}, E$ & $0.40\pm0.03\pm0.10\pm0.02$ &$ 0.54\pm0.04$ \\
  
  $D^0 \to K^- ( {K}^{*+}\to)K^+ \pi^0 $ &$T_{PV}, E$ & $1.53\pm0.06\pm0.35\pm0.15$ &$ 1.52\pm0.07$ \\
  
  $D^0 \to K_S^0 (\bar{K}^{*0}\to)K^- \pi^+ $ &$E$ & $0.12\pm0.03\pm0.00\pm0.02$ &$ 0.08 \pm 0.02$   \\ 
  
  $D^0 \to \bar K_S^0 ({K}^{*0}\to)K^+ \pi^- $ &$E$ & $0.12\pm0.03\pm0.00\pm0.02$ &$ 0.11\pm0.02$   \\ 
  
  $D^+ \to K^+ (\bar{K}^{*0}\to)K^- \pi^+ $ &$T_{VP}, A$ & $2.60\pm0.13\pm0.62\pm0.08$ &$ 2.49 ^{+0.08}_{-0.13}$   \\   
  
  $D^+ \to \bar K^0 ({K}^{*+}\to)K^+ \pi^0 $ &$T_{PV}, A$ & $4.26\pm0.12\pm0.94\pm0.42$ &$ $   \\   
  
  $D_s^+ \to \pi^+ ({K}^{*0}\to)K^+ \pi^- $ &$T_{VP}, A$ & $1.56\pm0.10\pm0.34\pm0.03$ &$1.40\pm0.24 $   \\    
 
  $D_s^+ \to \pi^0 ({K}^{*+}\to) K^+ \pi^0 $ &$C_{VP}, A$ & $0.16\pm0.02\pm0.03\pm0.00$ &$ $   \\    
  
  $D_s^+ \to \eta ({K}^{*+}\to)K^+ \pi^0 $ &$T_{PV}, C_{VP}, A$ & $0.47\pm0.08\pm0.16\pm0.08$ &$ $   \\ 

$D_s^+ \to \eta^\prime ({K}^{*+}\to)K^+ \pi^0 $ &$T_{PV}, C_{VP}, A$ & $0.19\pm0.01\pm0.05\pm0.03$ &$ $   \\ 
   
 \hline
   DCS &$V_{cd}^*V_{us}\,$ &&\\

$D^0 \to \pi^- ( {K}^{*+}\to)K_S^0 \pi^+ $ &$T_{VP}, E $ & $0.15\pm0.01\pm0.03\pm0.02$ &$0.11^{+0.06}_{-0.03} $ \\
 
$D^0 \to \pi^0 ({K}^{*0}\to)K^+ \pi^- $ &$C_{PV}, E$ & $0.06\pm0.01\pm0.01\pm0.01$ &$ $   \\ 

$D^0 \to \eta ({K}^{*0}\to)K^+ \pi^- $ &$C_{PV}, E$ & $0.02\pm0.00\pm0.00\pm0.00$ &$ $   \\ 

$D^+ \to \pi^0 ({K}^{*+}\to)K^+ \pi^0 $ &$T_{PV}, A$ & $0.14\pm0.00\pm0.03\pm0.01$ &$ $   \\     

$D^+ \to \pi^+ ({K}^{*0}\to)K^+ \pi^- $ &$C_{PV}, A$ & $0.21\pm0.01\pm0.05\pm0.02$ &$ 0.23\pm0.04$   \\   
  
$D^+ \to \eta ({K}^{*+}\to) K^+ \pi^0 $ &$T_{PV}, A$ & $0.06\pm0.00\pm0.01\pm0.01$ &$ $   \\ 

$D_s^+ \to K^+ ({K}^{*0}\to)K^+ \pi^- $ &$T_{VP}, C_{PV}$ & $0.02\pm0.00\pm0.00\pm0.00$ &$ 0.06\pm0.03$   \\   
   
 $D_s^+ \to  K^0 ({K}^{*+}\to) K^+ \pi^0 $ &$T_{PV}, C_{VP}$ & $0.06\pm0.00\pm0.02\pm0.01$ &$ $   \\   
                                                                            
 \hline \hline
\end{tabular}
\end{center}
\end{table}

The branching fractions of 
$D_{(s)} \to P_1\, ( \rho \to)\,  \pi \pi$ and  $D_{(s)} \to P_1\, (K^{*} \to) \, K \pi$
are listed in Tables~\ref{Prho} and~\ref{PKstar},respectively. 
In Table~\ref{PKstar}, we specifically list quasi-two-body decays 
$D_{(s)} \to P_1 K^{*} \to P_1 K \pi$ mediated by the strong decays $K^{*0} \to  K^+ \pi^-$, 
$ \bar{K}^{*0} \to  K^- \pi^+$, and $ K^{*-} \to  K^- \pi^0$,  
where the $K^*$ resonances involve contributions from $u \bar u$ sea quark pairs.  
For decays proceeding via $d \bar d$ sea quark pairs, such as 
$ K^{*0} \to  K^0 \pi^0$, $ \bar{K}^{*0} \to  \bar{K}^0 \pi^0$ and $ K^{*-} \to  \bar{K}^0 \pi^-$,
the corresponding branching ratios for $D_{(s)} \to P_1 K^{*} \to P_1 K \pi$ 
can be derived using the narrow-width approximation and 
are calculated by applying isospin conservation in
 the strong interaction dynamics of  $K^* \to K \pi $ decay.
The relations between branching ratios are explicitly expressed as
 \begin{eqnarray}
 \mathcal{B}(\bar{K}^{*0} \to K^- \pi^+) &=& 2\, \mathcal{B} (\bar{K}^{*0} \to \bar{K}^0 \pi^0),\\
 \mathcal{B} (K^{*-} \to \bar{K}^0 \pi^-)&=& 2\, \mathcal{B}(K^{*-} \to K^- \pi^0).
 \end{eqnarray}
 
The observed hierarchies in the branching fractions of 
$D_{(s)} \to P_1\, ( \rho \to)\,  \pi \pi $ and 
$D_{(s)} \to P_1\, (K^{*} \to) \, K \pi$ decays arise from 
two key factors: one is the CKM hierarchy.
CF decays ($V_{cs}^*V_{ud}$)  
dominate due to large CKM elements 
($|V_{cs}| \sim 0.97,|V_{ud}| \sim 0.97 $), leading to
branching fractions about $10^{-3} - 10^{-2} $.
SCS decays ($V_{cd(s)}^*V_{ud(s)}$)  
are suppressed by $|V_{cd}/V_{cs}|^2 \sim \lambda^2\sim 0.05$,
reducing branching fractions to $10^{-4} - 10^{-3} $.
DCS decays ($V_{cd}^*V_{us}$)  
suffer an additional suppression $|V_{us}/V_{ud}|^2 \sim \lambda^2$,
resulting in $\mathcal{B}\sim 10^{-5} - 10^{-4} $.
Another is topological diagram hierarchy. The relative strengths of 
diagrams follow $T_{VP/PV}>C_{VP/PV}>E>A $.
This effect explains the observed hierarchies within the same 
type of CKM transition (e.g., CF, SCS, or DCS).
For example, the branching fraction of $D_s^+ \to \pi^+ ({K}^{*0}\to)K^+ \pi^- $ 
(SCS, $T_{VP}$ dominant) is larger than $D_s^+ \to \pi^0 ({K}^{*+}\to) K^+ \pi^0 $
(SCS, $C_{VP}$ dominant).

Nearly all CF and SCS modes of $D_{(s)} \to P_1\, (K^{*} \to) \, K \pi$, 
as well as several DCS modes,
have already been experimentally measured,
as indicated in the last column of Table~\ref{PKstar} from PDG~\cite{ParticleDataGroup:2024cfk}.
To compare with experimental results of some modes, we apply the approximate relation 
$\Gamma (\bar K^0)\, =\, 2 \Gamma (\bar K_S^0)\,$ to convert $\bar K^0$ 
into the experimentally measurable $K_S^0$ component.
Our theoretical predictions for $D_{(s)} \to P_1\, (K^{*} \to) \, K \pi$ 
agree well with experimental data.
Other modes with comparable branching ratio, such as 
SCS modes $D^+ \to \bar K^0 ({K}^{*+}\to)K^+ \pi^0 $, $D^+ \to \bar K^0 ({K}^{*+}\to)K^+ \pi^0 $
and CF and SCS $D_s$ decays, are also experimentally accessible.

For the branching fractions of $D_{(s)} \to P_1\, ( \rho \to)\,  \pi \pi$,
we list the results of measured branching ratios of two-body decays of $D_{(s)} \to P_1\, \rho$ 
for comparison, as $\mathcal{B}(\rho \to \pi \pi ) \sim 100\%$.
Most of our results for quasi-two-body decay $D_{(s)} \to P_1 \rho \to P_1 \pi \pi$
are slightly smaller than those of the direct two-body decays 
$D_{(s)} \to P_1 \rho$. The difference can be attributed to 
the relatively broad resonance width $\Gamma_\rho$ of the $\rho(770)$ meson,
which invalidates the narrow-width approximation typically used to factorize decay amplitudes.
The $\rho(770)$ resonance dominates 
the $D^0 \to \pi^+  \pi^- \pi^0 $ decays,
as evidenced by BABAR's Dalitz analysis, where the fit fractions of 
$D^0 \to \pi^+ (\rho^-\to)\pi^- \pi^0$, $D^0 \to \pi^- (\rho^+\to)\pi^+ \pi^0$, 
and $D^0 \to \pi^0 (\rho^0 \to)\pi^+ \pi^-$ 
are $34.6 \pm0.8\pm 0.3$, $67.8 \pm0.0\pm 0.6$, and $26.2 \pm0.5\pm 1.1$, 
respectively~\cite{BaBar:2007dro}. 
Our results in the FAT approach, incorporating updated 
experimental data, yield smaller branching ratios than those derived from 
 topological diagram methods~\cite{Cheng:2021yrn}, yet show closer agreement 
 with the corresponding two-body decays of  $D^0 \to \pi^+ \rho^-$,  
 $D^0 \to \pi^- \rho^+$ and $D^0 \to \pi^0 \rho^0$.
 This improved consistency suggests that the FAT framework 
 better accounts for nonfactorizable effects
 (e.g., final-state interactions and resonance interference) 
 and flavor $SU(3)$ asymmetry in the quasi-two-body decay formalism.
 The latter arises from mass differences between $u, d$, and $s$
quarks and dynamical asymmetries in decay amplitudes, 
which are not fully captured by topological diagram approaches 
relying on exact SU(3) symmetry assumptions.

As shown in Tables~\ref{Prho} and~\ref{PKstar}, the $D_{(s)} \to K \pi \pi $ decays 
can proceed via contributions from both $\rho$ and $K^*$ resonances. 
For example, $D^0 \to K^+ \pi^- \pi^0$ receives contributions from 
$\rho^-$, $K^{*+}$, and $K^{*0}$ resonances.
$D^+ \to K^+ \pi^+ \pi^-$ is mediated by  
$\rho^0$ and $K^{*0}$ resonances.
$D_s^+ \to K^+ \pi^+ \pi^-$ similarly involves 
$\rho^0$ and $K^{*0}$ resonances.
In this work, we focus on calculating 
the branching ratios for these decay modes.
The analysis of {\it CP} asymmetries, arising from 
interference effects between overlapping resonances
(e.g., $\rho-K^*$ interference in phase space), will
be addressed in future work.
Such interference phenomena are critical for understanding dynamics 
in three-body decays but require a detailed amplitude analysis of the Dalitz plot.


\subsection{Branching fractions of $D_{(s)} \to P_1 \phi \to P_1 K\bar K$}
\begin{table}[tbhp]
\caption{Same as Table \ref{Prho} for quasi-two-body decays 
$D_{(s)} \to P_1 (\phi \to ) K K$  in units of $10^{-3}$.}
 \label{Pphi} 
\begin{center}
\begin{tabular}{cccc}
 \hline \hline
Decay modes     &  Amplitudes & FAT  results       &  Experiment   \\
\hline
CF & $V_{cs}^*V_{ud}\, $ & &\\

$D^{0}\to K_S^{0}(\phi \to)K^{+}K^{-}$ &$ E $ & $1.62\pm0.16\pm0\pm0.23$ &$2.03\pm 0.15 $   \\ 

$\phantom{D}\to\bar{K}^{0}(\phi \to)K^{0}\bar{K}^{0}$ &$ $ & $2.20\pm0.22\pm0\pm0.31$ &$ $   \\     

$D^{+}_{s}\to\pi^{+}(\phi \to)K^{+}K^{-}$ &$T_{VP} $ &$23.94\pm0.14\pm1.53\pm0.02$  &$ 22.2 \pm 0.6$   \\

$\phantom{D}\to\pi^{+}(\phi \to)K^{0}\bar{K}^{0}$ &$ $ &  $16.31\pm0.10\pm1.16\pm0.02$&$ $   \\ 

\hline     
SCS &$V_{c d(s)}^*V_{ud(s)}\, $ &&\\
$D^{0}\to \pi^{0}(\phi \to)K^{+}K^{-}$ &$C_{PV} $ & $0.62 \pm0.01\pm0.07\pm0.03 $ &$ 0.66 \pm 0.04$ \\
 
$\quad \to \pi^{0}(\phi \to)K^{0}\bar{K}^{0} $ &$ $ & $0.43 \pm0.01\pm0.04\pm0.02$ &$ $ \\

$D^{0}\to\eta(\phi \to)K^{+}K^{-}$ &$ C_{PV}, E$ & $0.13\pm0.03\pm0\pm0.02$ &$ $\\

$\phantom{D}\to\eta(\phi \to)K^{0}\bar{K}^{0}$ &$ $ & $0.09\pm0.02\pm0\pm0.01$ &$ $   \\       
 
$D^{+}\to\pi^{+}(\phi \to)K^{+}K^{-} $ &$C_{PV} $ &$3.18\pm0.06\pm0.34\pm0.17$  &$ 2.69 ^{+0.07}_{-0.08}$   \\ 

$\phantom{D‘} \to\pi^{+}(\phi \to)K^{0}\bar{K}^{0} $ &$ $ & $2.18\pm0.03\pm0.18\pm0.09$ &$ $   \\                

$D^{+}_{s}\to K^{+}(\phi \to)K^{+}K^{-}$ &$T_{VP}, C_{PV}, A $ & $0.13\pm0.03\pm0.06\pm0.01$  &$ 0.088\pm0.02$   \\    
 
$\phantom{D}\to K^{+}(\phi \to)K^{0}\bar{K}^{0}$ &$ $ &$0.09\pm0.02\pm0.04\pm0.01$ &$ $   \\    
 
\hline
 DCS &$V_{cd}^*V_{us}\,$ &&\\
$D^{0}\to K^{0}(\phi \to)K^{+}K^{-}$ &$ E $ & $0.009\pm0.001\pm0\pm0.001$ &$ $   \\

$\phantom{D}\to K^{0}(\phi \to)K^{0}\bar{K}^{0}$ &$ $ & $0.006\pm0.001\pm0\pm0.001$ &$ $   \\

$D^{+}\to K^{+}(\phi \to)K^{+}K^{-} $ &$ A $ & $ (3.7 \pm 0.6\pm 0\pm 0.5)\times 10^{-3}$ &$ (4.4 \pm 0.6)\times 10^{-3}$   \\  

$\phantom{D}\to K^{+}(\phi \to)K^{0}\bar{K}^{0}$ &$ $ & $(2.5 \pm 0.4\pm 0\pm 0.4)\times 10^{-3}$ &$ $   \\   

 \hline \hline
\end{tabular}
\end{center}
\end{table}
We present the branching fraction results of $D_{(s)} \to P_1 ( \phi \to) K\bar K$ decays in Table~\ref{Pphi} 
alongside corresponding PDG values~\cite{ParticleDataGroup:2024cfk} in the last column.
The branching ratios of $D_{(s)} \to \pi, K ( \phi \to) K\bar K$ 
have been measured by experiments and the results in the FAT approach 
agree well with these experimental data. 
In addition to listing the quasi-two-body decay modes involving $u \bar u$ sea quarks in $\phi \to K^+K^-$,
we also include those with $d \bar d$ sea quarks in $\phi \to K^0 \bar K^0$.
By summing contributions from $D_{(s)} \to  P_1 (\phi \,\to) K^+K^- $
and $D_{(s)} \to  P_1 (\phi \,\to) K^0 \bar{K}^0$ decays (which share the same weak transition 
but different strong decay), we recover the the corresponding two-body decay $D \to  P_1 \phi \,$ 
branching fractions through the narrow-width approximation.
For instance, the sum of our predictions for the unmeasured 
quasi-two-body decays $D^0 \to \eta( \phi \to) K^+  K^- $ and 
$D^0 \to \eta( \phi \to) K^0 \bar K^0 $ 
is in agreement with the PDG value for 
the two-body decay 
$\mathcal{B}(D^0 \to  \eta \phi )=(0.184 \pm 0.012) \times 10^{-4}$.
  
\subsection{The virtual effects of $D_{(s)} \to P_1(\omega \to)K\bar K$}\label{subBWT}
\begin{table}[tbhp]
\caption{The virtual effects of $D_{(s)} \to P_1 (\omega \to)K \bar K$ decays,
 which happen when the pole masses of $ \omega $ are smaller than 
 the invariant mass of $K \bar K$.}
 \label{Pomega}
\begin{center}
\begin{tabular}{ccc}
 \hline \hline
Decay modes     &  Amplitudes & FAT  results         \\
\hline
CF & $V_{cs}^*V_{ud}\, $ & \\

$D^{0}\to\bar{K}^{0}(\omega \to)K^{+}K^{-}$ &$ C_{VP}, E$ & $(0.90\pm0.07\pm0.13\pm0.03)\times 10^{-5}$   \\ 

$D^{+}_{s}\to\pi^{+}(\omega \to)K^{+}K^{-}$ &$ A$ & $(3.28\pm0.22\pm0\pm0.43)\times 10^{-6}$   \\

\hline                    
 SCS &$V_{c d(s)}^*V_{ud(s)}\, $ &\\
 
$D^{0}\to \pi^{0}(\omega \to)K^{+}K^{-}$ &$C_{PV}, C_{VP}, E $ & $(7.11\pm1.42\pm1.99\pm1.11)\times 10^{-7}$  \\

$D^{0}\to\eta(\omega \to)K^{+}K^{-}$ &$ C_{PV},C_{VP}, E$ & $(1.05\pm0.06\pm0.13\pm0.05)\times 10^{-6}$ \\

$D^{+}\to\pi^{+}(\omega \to)K^{+}K^{-} $ &$T_{VP}, C_{PV}, A $ & $(5.49\pm0.84\pm2.65\pm1.07)\times 10^{-7}$   \\ 

$D^{+}_{s}\to K^{+}(\omega \to)K^{+}K^{-}$ &$ C_{PV},A$ & $(1.86\pm0.09\pm0.35\pm0.19)\times 10^{-6}$  \\   
  
\hline
 DCS &$V_{cd}^*V_{us}\,$ &\\

$D^{0}\to K^{0}(\omega \to)K^{+}K^{-}$ &$ C_{VP}, E$ & $(2.56\pm0.20\pm0.38\pm0.09)\times 10^{-8}$   \\

$D^{+}\to K^{+}(\omega \to)K^{+}K^{-} $ &$ T_{VP}, A$ & $(9.69\pm0.31\pm1.82\pm0.11)\times 10^{-8}$    \\  

 \hline \hline
\end{tabular}
\end{center}
\end{table}
The decay modes $D_{(s)} \to P_1 (\omega \to ) K \bar K$ denote a distinct 
category of processes where the $ \omega$ resonance pole mass ($m_\omega  \approx 783 $ MeV)
lies below the $ K \bar K$ production threshold 
( $\sim 987 $ MeV for $K^+ K^-$ and $\sim 994 $ MeV for
  $K^0 \bar K^0$ ).
  These decays proceed via virtual transitions [the Breit-Wigner-tail (BWT) effect],
  a mechanism previously studied in $B$ meson decays
   on the theoretical side~\cite{Zhou:2021yys,Zhou:2023lbc,Zhou:2024qmm} and 
   by experiments in Belle~\cite{Belle:2006wbx} and LHCb~\cite{LHCb:2016lxy}.
   The BWT effects play an important role in probing  
the shape function of vector resonances 
and analyzing the dynamics of three-body decays mechanisms.
In Table~\ref{Pomega}, we quantify these virtual effects using the FAT approach.
Predicted branching fractions for most modes lie in the range $10^{-7}-10^{-6}$,
rendering them experimentally inaccessible with current precision. However, 
CF modes approaching $10^{-5}$ may warrant experimental analysis. 
We exclusively consider the subprocesses $ \omega \to K^+ K^-$ in our analysis.
The omission of  $\omega \to K^0 \bar K^0$ is justified by  
the negligible phase space difference between $K^+ K^-$ and $K^0 \bar{K}^0$
final states (stemming from the small $K^0- K^+$ mass difference, $\Delta_{m_K}  \approx 4$ MeV),
which results in virtually identical BWT behaviors.
The Breit-Wigner propagators in Eq.~(\ref{RBW}) exhibit specific kinematics 
in these decays: For the invariant mass square $s \geq 1\, \mathrm{GeV}^2 $ 
(for above the $\omega$ pole), the imaginary term  
$\mathrm{i}\, m_{\omega} \Gamma_{\omega}(s)$ 
becomes subdominant.
Consequently, virtual contributions show minimal sensitivity to
the $\omega$ meson's full decay width $\Gamma_{\omega}(s)$,
as the propagator behavior is governed by the $(m_\omega^2-s)$
denominator term.


 \section{Conclusion}\label{sec:4}
Theoretical studies of charm meson decays remain limited 
due to the intermediate mass scale of the charm meson,
which lies between perturbative and nonperturbative regimes.
This gap is particularly pronounced for three-body $D$-meson
decays, despite recent measurements of such decays with 
resonance contributions by BESIII and other experiments.
In this work, we systematically investigate the  quasi-two-body 
decays $D_{(s)} \to P_1 (V \to)  P_2 P_3 $, where $V$ denotes
intermediate vector resonances ($\rho, K^*, \omega$, $\phi$),
decaying strongly to pseudoscalars $P_2$ and $P_3$.
These processes originate from  $c \rightarrow \,d/s $
transitions forming $P_1 V$ intermediate states.
Using the FAT approach, we can describe well the 
nonperturbative contributions for the intermediate
two-body decays $D_{(s)}  \to P_1 V$.
The subsequent resonant dynamics are modeled 
via the relativistic RBW line shape,
consistent with experimental analyses. 
We categorize the decays into four types based on the vector resonance:
$D_{(s)} \to P_1 (\rho \to)  \pi \pi $, 
$D_{(s)} \to P_1 (K^* \to) K \pi $, $D_{(s)} \to P_1 (\phi \to) K K $,
and $D_{(s)} \to P_1 (\omega \to) K K $. 
The first three classes are governed by pole dynamics, 
while the $\omega$-mediated channel arises predominantly 
from the BWT interference effect.
This systematic framework bridges theoretical predictions with 
experimental resonance analyses in three-body $D$-meson decays.

In this work, we calculate the branching ratios for the four categories of 
quasi-two-body decay modes.
Utilizing updated nonperturbative parameters derived from a global fit
to precise experimental data, our results are in alignment with 
existing measurements. 
The theoretical uncertainties in our approach are small, 
owing to the precise determination of nonfactorizable parameters
from data-driven analyses.
For unmeasured modes, we predict their branching ratios and
are waiting for future experiments to test,
especially for those on the order of $10^{-4}-10^{-3}$, 
which are promising candidates for future observation in 
high-precision experiments like BESIII and LHCb.
Notably, the fourth category decay $D_{(s)} \to P_1 (\omega \to) K K $
proceeds via the tail effects of the $\omega$ resonance in the $KK$
channel. This mechanism provides an opportunity to probe 
the shape function of vector resonances, offering insights into 
the interplay between resonance dynamics and 
three-body decay mechanisms.


\section*{Acknowledgments}
The work is supported by the National Natural Science Foundation of China
under Grant No.12465017.

\bibliographystyle{bibstyle}
\bibliography{refs}

\providecommand{\href}[2]{#2}\begingroup\raggedright\begin{thebibliography}{100}

\bibitem{BESIII:2016nrs}
M.~Ablikim \textit{et al.},  (BESIII Collaboration), { {Measurements of the
  branching fractions for $D^+\to K^0_S K^0_SK^+$, $K^0_SK^0_S\pi^+$ and $D^0\to
  K^0_SK^0_S$, $K^0_SK^0_SK^0_S$}},  {Phys. Lett. B} {\bf 765}, 231 (2017).

\bibitem{BESIII:2018hui}
 M.~Ablikim \textit{et al.},   (BESIII Collaboration), {{Measurement of singly
  Cabibbo-suppressed decays $D^{0}\to\pi^{0}\pi^{0}\pi^{0}$,
  $\pi^{0}\pi^{0}\eta$, $\pi^{0}\eta\eta$ and $\eta\eta\eta$}},  {Phys.
  Lett. B} {\bf 781}, 368 (2018).

\bibitem{BESIII:2018exp}
M.~Ablikim \textit{et al.},  (BESIII Collaboration), {{Observation of
  $D^{0(+)}\to K^0_S\pi^{0(+)}\eta^\prime$ and improved measurement of $D^0\to
  K^-\pi^+\eta^\prime$}},  { Phys. Rev. D} {\bf 98}, 092009 (2018).

\bibitem{BESIII:2018pku}
 M.~Ablikim \textit{et al.},  (BESIII Collaboration), {{Measurements of the
  absolute branching fractions and $CP$ asymmetries for $D^+\rightarrow
  K_{S,L}^0 K^+(\pi^0)$}},  {Phys. Rev. D} {\bf 99}, 032002 (2019).

\bibitem{BESIII:2019xhl}
 M.~Ablikim \textit{et al.},  (BESIII Collaboration),{ {Observation of
  $D^+\to\eta\eta\pi^+$ and improved measurement of
  $D^{0(+)}\to\eta\pi^+\pi^{-(0)}$}},  {Phys. Rev. D} {\bf 101}, 052009 (2020).

\bibitem{BESIII:2021jnf}
M.~Ablikim \textit{et al.},  (BESIII Collaboration), { {Amplitude analysis of $D_s^+ \to \pi^+ \pi^- \pi^+$
  }}, {Phys. Rev. D} {\bf 106}, 112006 (2022).

\bibitem{LHCb:2011nqf}
R.~Aaij \textit{et al.},  (LHCb Collaboration), { {Search for CP violation in
  $D^{+} \to K^{-}K^{+}\pi^{+}$ decays}},  {Phys. Rev. D} {\bf 84}, 112008 (2011)
 
\bibitem{LHCb:2013qzm}
R.~Aaij \textit{et al.},  (LHCb Collaboration), { {Search for CP violation in the
  decay $D^+ \to \pi^-\pi^+\pi^+$}},  {Phys. Lett. B} {\bf 728}, 585 (2014).

\bibitem{LHCb:2014nnj}
R.~Aaij \textit{et al.},  (LHCb Collaboration), {{Search for CP violation in $D^0
  \to \pi^- \pi^+ \pi^0$ decays with the energy test}},  {Phys. Lett. B}
  {\bf 740} ,158 (2015). 

\bibitem{LHCb:2023qne}
R.~Aaij \textit{et al.}, (LHCb Collaboration), { {Search for $\it{CP}$ violation
  in $D_{(s)}^{+}\rightarrow K^{-}K^{+}K^{+}$ decays}},  {JHEP} {\bf 07}
  (2023) 067.

\bibitem{LHCb:2024rkp}
R.~Aaij \textit{et al.},  (LHCb Collaboration), { {Measurement of CP Violation
  Observables in $D^+ \to K^- K^+ \pi^+$ Decays}},  
  { Phys. Rev. Lett.} {\bf 133}, 251801, (2024).

\bibitem{CLEO:2000fvk}
S.~Kopp \textit{et al.},  (CLEO Collaboration),  {{Dalitz analysis of the decay $D^0 \to K^- \pi^+ \pi^0 $}},  
{ Phys. Rev. D} {\bf 63}, 092001 (2001).

\bibitem{CLEO:2007rrw}
S.~Dobbs \textit{et al.},  (CLEO Collaboration), {{Measurement of absolute
  hadronic branching fractions of D mesons and $e^+ e^- \to D \bar{D}$
  cross-sections at the $\psi(3770)$}},  {Phys. Rev. D} {\bf 76}, 112001 (2007).

\bibitem{CLEO:2008msk}
P.~Rubin \textit{et al.},  (CLEO Collaboration), {{Search for CP violation in the
  Dalitz-Plot analysis of $D^\pm \to K^+ K^- \pi^\pm$ }},  { Phys. Rev.
  D} {\bf 78}, 072003 (2008).

\bibitem{CLEO:2010enr}
R.~A. Briere \textit{et al.},  (CLEO Collaboration), {{Analysis of $D^+ \to K^-
  \pi^+ e^+ \nu_e$ and $D^+ \to K^- \pi^+ \mu^+ \nu_\mu$ semileptonic decays}},
   {Phys. Rev. D} {\bf 81}, 112001 (2010).

\bibitem{CLEO:2011cnt}
 N.~Lowrey \textit{et al.},  (CLEO Collaboration), {{Analysis of the decay $D^0
  \to K^0_S \pi^0 \pi^0$}},  { Phys. Rev. D} {\bf 84}, 092005 (2011).

\bibitem{CLEO:2012obf}
J.~Insler \textit{et al.},  (CLEO Collaboration), { {Studies of the decays $D^0
  \rightarrow K_S^0K^-\pi^+$ and $D^0 \rightarrow K_S^0K^+\pi^-$}},  { Phys.
  Rev. D} {\bf 85}, 092016 (2012).

\bibitem{CLEO:2013bae}
P.~U.~E. Onyisi \textit{et al.},  (CLEO Collaboration), {{Improved measurement of
  absolute hadronic branching fractions of the $D_s^+$ meson}},  { Phys.
  Rev. D} {\bf 88},032009 (2013).

\bibitem{CLEO:2013rjc}
G.~Bonvicini \textit{et al.},  (CLEO Collaboration), {{Updated measurements of
  absolute $D^+$ and $D^0$ hadronic branching fractions and $\sigma(e^+e^-\to
  D\overline{D})$ at $E_\mathrm{cm} = 3774$ MeV}},  { Phys. Rev. D} {\bf 89},
  072002 (2014).

\bibitem{Belle:2008qrk}
K.~Arinstein \textit{et al.},  (Belle Collaboration), { {Measurement of the ratio
  $B(D^0 \to \pi^{+} \pi^{-} \pi^0)$ / $B(D^0 \to K^{-} \pi^{+} \pi^0)$ and the
  time-integrated CP asymmetry in $D^0 \to \pi^{+} \pi^{-} \pi^0$}},  {
  Phys. Lett. B} {\bf 662}, 102 (2008).

\bibitem{Belle-II:2024gfc}
I.~Adachi \textit{et al.},  (Belle-II, Belle Collaboration),  {{Model-independent
  measurement of $D^0$-$\overline{D}{}^0$ mixing parameters in $D^0\rightarrow
  K^0_{S}\pi^+\pi^-$ decays at Belle and Belle II}},
 arXiv:2410.22961

\bibitem{BaBar:2005rek}
B.~Aubert \textit{et al.},  (BaBar Collaboration), { {Measurement of $\gamma$ in
  $B^\mp \to D^{(*)} K^\mp$ decays with a Dalitz analysis of $D \to K^0_S \pi^-
  \pi^+$}},  { Phys. Rev. Lett.} {\bf 95}, 121802 (2005).

\bibitem{BaBar:2006dxf}
B.~Aubert \textit{et al.},  (BaBar Collaboration),{ {Precise branching ratio
  measurements of the decays $D^0 \to \pi^{-} \pi^{+} \pi^0$ and $D^0 \to K^{-}
  K^{+} \pi^0$}},  { Phys. Rev. D} {\bf 74} 091102 (2006) 

\bibitem{BaBar:2008xzl}
 B.~Aubert \textit{et al.},  (BaBar Collaboration), { {Search for CP violation in
  neutral D meson Cabibbo-suppressed three-body decays}},  { Phys. Rev. D}
  {\bf 78}, 051102 (2008).
  
\bibitem{BaBar:2008nlp}
B.~Aubert \textit{et al.},  (BaBar Collaboration), { {Dalitz plot analysis of $D_s^
  + \to \pi^{+} \pi^{-} \pi^{+}$}},  {Phys. Rev. D} {\bf 79}, 032003 (2009).

\bibitem{BaBar:2012ize}
J.~P. Lees \textit{et al.},  (BaBar Collaboration), {{Search for direct CP
  violation in singly Cabibbo-suppressed
  $D^{\pm} \to K^+ K^- \pi^\pm$
  decays}},  {Phys. Rev. D} {\bf 87} , 52010 (2013).

\bibitem{Buras:1985xv}
A.~J. Buras, J.~M. Gerard, and R.~Ruckl, { {1/n expansion for exclusive and
  inclusive charm decays}},  { Nucl. Phys. B} {\bf 268}, 16 (1986) .

\bibitem{BaBar:2007soq}
B.~Aubert \textit{et al.},  (BaBar Collaboration), { {Amplitude analysis of the
  decay $D^0 \to K^- K^+ \pi^0$ }},  { Phys. Rev. D} {\bf 76}, 011102 (2007).

\bibitem{Beneke:2000ry}
M.~Beneke, G.~Buchalla, M.~Neubert, and C.~T. Sachrajda, { {QCD
  factorization for exclusive, nonleptonic B meson decays: General arguments
  and the case of heavy light final states}},  { Nucl. Phys. B} {\bf 591}, 313
  (2000).

\bibitem{Lu:2000em}
C.-D. Lu, K.~Ukai, and M.-Z. Yang, {{Branching ratio and CP violation of $B \to \pi \pi $ 
decays in perturbative QCD approach}},  { Phys.
  Rev. D} {\bf 63}, 074009 (2001).

\bibitem{Keum:2000wi}
Y.~Y. Keum, H.-N. Li, and A.~I. Sanda, { {Penguin enhancement and $B \to K
  \pi$ decays in perturbative QCD}},  { Phys. Rev. D} {\bf 63}, 054008 (2001).

\bibitem{Ryd:2009uf}
A.~Ryd and A.~A. Petrov, {{Hadronic $D$ and $D_s$ meson decays}},  { Rev.
  Mod. Phys.} {\bf 84}, 65 (2012).

\bibitem{Bhattacharya:2008ke}
B.~Bhattacharya and J.~L. Rosner, { {Decays of charmed mesons to PV final
  states}},  { Phys. Rev. D} {\bf 79}, 034016 (2009).

\bibitem{Wang:2021ail}
W.-F. Wang, { {Subprocesses
 ${\rho}(770,1450)\to K \, K^-$ for the
  three-body hadronic D meson decays}},  {Phys. Rev. D} {\bf 104} 116019  (2021).

\bibitem{Brod:2011re}
J.~Brod, A.~L. Kagan, and J.~Zupan, { {Size of direct CP violation in singly
  Cabibbo-suppressed D decays}},  { Phys. Rev. D} {\bf 86}, 014023 (2012).
  
\bibitem{Cheng:2010ry}
H.-Y. Cheng and C.-W. Chiang, {{Two-body hadronic charmed meson decays}},
  { Phys. Rev. D} {\bf 81}, 074021 (2010).

\bibitem{Cheng:2012wr}
H.-Y. Cheng and C.-W. Chiang, { {Direct CP violation in two-body hadronic
  charmed meson decays}},  { Phys. Rev. D} {\bf 85}, 034036 (2012).

\bibitem{Cheng:2012xb}
H.-Y. Cheng and C.-W. Chiang, { {SU(3) symmetry breaking and CP violation in
  D -\ensuremath{>} PP decays}},  { Phys. Rev. D} {\bf 86}, 014014 (2012).
  
\bibitem{Chen:2012am}
C.-H. Chen, C.-Q. Geng, and W.~Wang, { {CP violation in $D^0 \to (K^- K^+,
  \pi^- \pi^+)$ from diquarks}},  { Phys. Rev. D} {\bf 85}, 077702 (2012).

\bibitem{Cheng:2016ejf}
H.-Y. Cheng, C.-W. Chiang, and A.-L. Kuo, {{Global analysis of two-body
  D$\to$VP decays within the framework of flavor symmetry}},  { Phys. Rev.
  D} {\bf 93}, 114010 (2016).
  
\bibitem{Cheng:2019ggx}
H.-Y. Cheng and C.-W. Chiang, { {Revisiting CP violation in $D\to P\!P$ and
  $V\!P$ decays}},  { Phys. Rev. D} {\bf 100}, 093002 (2019).

\bibitem{Hsiao:2019ait}
Y.-K. Hsiao, Y.~Yu, and B.-C. Ke, { {Resonant $a_0(980)$ state in triangle
  rescattering $D_s^+\rightarrow \pi ^+\pi ^0\eta $ decays}},  { Eur. Phys.
  J. C} {\bf 80}, 895 (2020).

\bibitem{Cheng:2021yrn}
H.-Y. Cheng and C.-W. Chiang, {{CP violation in quasi-two-body
  D\textrightarrow{}VP decays and three-body D decays mediated by vector
  resonances}},  { Phys. Rev. D} {\bf 104}, 073003 (2021).

\bibitem{Li:2012cfa}
H.-n. Li, C.-D. Lu, and F.-S. Yu, { {Branching ratios and direct CP
  asymmetries in $D\to PP$ decays}},  { Phys. Rev. D} {\bf 86}, 036012 (2012).

\bibitem{Li:2013xsa}
Q.~Qin, H.-n. Li, C.-D. L\"u, and F.-S. Yu, { {Branching ratios and direct
  CP asymmetries in $D\to PV$ decays}},  { Phys. Rev. D} {\bf 89}, 054006 (2014).
  
\bibitem{Qin:2021tve}
Q.~Qin, C.~Wang, D.~Wang, and S.-H. Zhou, { {The factorization-assisted
  topological-amplitude approach and its applications}},  { Front. Phys.
  (Beijing)} {\bf 18}, 64602 (2023).

\bibitem{Zhou:2018suj}
H.~Zhou, B.~Zheng, and Z.-H. Zhang, {{Analysis of $CP$ violation in $ D^0
  \to K^+ K^- \pi^0 $}},  { Adv. High Energy Phys.}, 7627308 (2018).

\bibitem{Wang:2025rkr}
W.-F. Wang, J.-Y. Xu, S.-H. Zhou, and P.-P. Shi, { {Contributions of
  $\rho(770,1450)\to \omega\pi$ for the Cabibbo-favored $D \to h\omega\pi$
  decays}}, arXiv:2502.11159

\bibitem{Dery:2021mll}
A.~Dery, Y.~Grossman, S.~Schacht, and A.~Soffer, { {Probing the $\Delta U=0$
  rule in three body charm decays}},  { JHEP} {\bf 05} (2021) 179.

\bibitem{Molina:2019udw}
R.~Molina, J.-J. Xie, W.-H. Liang, L.-S. Geng, and E.~Oset, {{Theoretical
  interpretation of the $D^+_s \to \pi^+ \pi^0 \eta$ decay and the nature of
  $a_0(980)$}},  { Phys. Lett. B} {\bf 803}, 135279 (2020).
  
\bibitem{Toledo:2020zxj}
G.~Toledo, N.~Ikeno, and E.~Oset, { {Theoretical study of the $D^0 \to K^-
  \pi^+ \eta$ reaction}},  { Eur. Phys. J. C} {\bf 81}, 268 (2021).

\bibitem{Roca:2020lyi}
L.~Roca and E.~Oset, { {Scalar resonances in the $D^+\to K^-K^+K^+$ decay}},
   { Phys. Rev. D} {\bf 103}, 034020 (2021).

\bibitem{Dai:2021owu}
L.~R. Dai, E.~Oset, and L.~S. Geng, { {The $D_s^+ \rightarrow \pi ^+ K_S^0
  K_S^0 $ reaction and the $I=1$ partner of the $f_0(1710)$ state}},  {Eur.
  Phys. J. C} {\bf 82}, 225 (2022).

\bibitem{Dai:2023jix}
L.~R. Dai and E.~Oset, { {Dynamical generation of the scalar $f_0(500)$,
  $f_0(980)$, and $K_0^*(700)$ resonances in the
  $D_s \to K^+ \pi^+ \pi^-$ reaction}},  {
  Phys. Rev. D} {\bf 109}, 054008 (2024).

\bibitem{Bayar:2023azy}
M.~Bayar, R.~Molina, E.~Oset, M.-Z. Liu, and L.-S. Geng, { {Subtleties in
  triangle loops for
  $D_s^+ \to \rho \eta\to \pi^+ \pi^0 \eta$
  in $a_0(980)$ production}},  { Phys. Rev. D} {\bf 109}, 076027 (2024).

\bibitem{Ikeno:2024fjr}
N.~Ikeno, J.~M. Dias, W.-H. Liang, and E.~Oset, {{$D^+ \rightarrow K_s^0
  \pi ^+ \eta $ reaction and $a_0(980)^+$}},  { Eur. Phys. J. C} {\bf 84}, 469
  (2024).

\bibitem{Ikeno:2021kzf}
N.~Ikeno, M.~Bayar, and E.~Oset, { {Combined theoretical study of the $D^+
  \to \pi^+ \eta \eta$ and $D^+ \to \pi^+ \pi^0 \eta $ reactions}},  { Eur.
  Phys. J. C} {\bf 81}, 377 (2021).

\bibitem{Boito:2017jav}
D.~Boito, J.~P. Dedonder, B.~El-Bennich, R.~Escribano, R.~Kaminski, L.~Lesniak,
  and B.~Loiseau, { {Parametrizations of three-body hadronic $B$- and
  $D$-decay amplitudes in terms of analytic and unitary meson-meson form
  factors}},  { Phys. Rev. D} {\bf 96}, 113003 (2017).

\bibitem{Guo:2018orw}
P.-F. Guo, D.~Wang, and F.-S. Yu, { {Strange axial-vector mesons in $D$
  meson decays}},  arXiv:1801.09582.

\bibitem{Yu:2021euw}
Y.~Yu, Y.-K. Hsiao, and B.-C. Ke, { {Study of the $D_s^+\rightarrow a_0(980)
  \rho $ and $a_0(980) \omega $ decays}},  { Eur. Phys. J. C} {\bf 81}, 1093
  (2021).

\bibitem{Song:2025lmj}
W.-J. Song, S.-Q. Wang, Q.~Qin, and Y.~Li, { {CP violation analysis of
  $D^0\rightarrow M^0 K \rightarrow M^0(\pi ^\pm \ell ^\mp \nu )$}},  { Eur.
  Phys. J. C} {\bf 85}, 300 (2025).

\bibitem{Niecknig:2015ija}
F.~Niecknig and B.~Kubis, {{Dispersion-theoretical analysis of the D$^{+}$
  \textrightarrow{} K$^{0}$ \ensuremath{\pi}$^{+}$ \ensuremath{\pi}$^{+}$
  Dalitz plot}},  { JHEP} {\bf 10} (2015) 142.

\bibitem{Nakamura:2015qga}
S.~X. Nakamura, { {Coupled-channel analysis of $D^+\to K^- \pi^+\pi^+$
  decay}},  {Phys. Rev. D} {\bf 93}, 014005 (2016).

\bibitem{Magalhaes:2011sh}
P.~C. Magalhaes, M.~R. Robilotta, K.~S. F.~F. Guimaraes, T.~Frederico,
  W.~de~Paula, I.~Bediaga, A.~C.~d. Reis, C.~M. Maekawa, and G.~R.~S.
  Zarnauskas, {{Towards three-body unitarity in $D^+ \to K^- \pi^+
  \pi^+$}},  { Phys. Rev. D} {\bf 84}, 094001 (2011).

\bibitem{Aoude:2018zty}
R.~T. Aoude, P.~C. Magalh\~aes, A.~C. Dos~Reis, and M.~R. Robilotta, {
  {Multimeson model for the $D^+\to K^+K^-K^+$ decay amplitude}},  { Phys.
  Rev. D} {\bf 98}, 056021 (2018).

\bibitem{Li:2020xrz}
H.-N. Li, H.~Umeeda, F.~Xu, and F.-S. Yu, { {$D$ meson mixing as an inverse
  problem}},  {Phys. Lett. B} {\bf 810}, 135802 (2020).

\bibitem{Saur:2020rgd}
M.~Saur and F.-S. Yu, { {Charm $CPV$: observation and prospects}},  {
  Sci. Bull.} {\bf 65}, 1428 (2020).

\bibitem{BESIII:2014oag}
 M.~Ablikim \textit{et al.},  (BESIII Collaboration), { {Amplitude Analysis of the
  $D^+ \to K^0_S \pi^+ \pi^0$ Dalitz Plot}},  {Phys. Rev. D} {\bf 89}, 052001
  (2014).

\bibitem{BESIII:2022vaf}
M.~Ablikim \textit{et al.},  (BESIII Collaboration), { {Amplitude analysis and
  branching fraction measurement of the decay $D_{s}^{+} \to K^+\pi^+\pi^-$}},
  { JHEP} {\bf 08} (2022) 196.
  
\bibitem{BESIII:2020hfw}
 M.~Ablikim \textit{et al.},  (BESIII Collaboration), { {Analysis of the decay
  $D^0\rightarrow K_{S}^{0} K^{+} K^{-}$}},
arXiv:2006.02800.

\bibitem{BESIII:2020ctr}
M.~Ablikim \textit{et al.},  (BESIII Collaboration), { {Amplitude analysis and
  branching fraction measurement of $D_{s}^{+} \rightarrow
  K^{+}K^{-}\pi^{+}$}},  { Phys. Rev. D} {\bf 104}, 012016 (2021).

\bibitem{BESIII:2021xox}
M.~Ablikim \textit{et al.}, (BESIII Collaboration), { {Amplitude analysis and
  branching-fraction measurement of $ {D}_s^{+}\to {K}_S^0{\pi}^{+}{\pi}^0 $}},
   {JHEP} {\bf 06} (2021) 181.

\bibitem{BESIII:2021dmo}
M.~Ablikim \textit{et al.}, (BESIII Collaboration), { {Study of the decay $D^+\to
  K^*(892)^+ K_S^0$ in $D^+\to K^+ K_S^0 \pi^0$}},  { Phys. Rev. D} {\bf
  104}, 012006 (2021).

\bibitem{BESIII:2021qsa}
M.~Ablikim \textit{et al.}, (BESIII Collaboration), { {Observation of the doubly
  Cabibbo-suppressed decays~$D^+\to K^+\pi^0\pi^0$ and $D^+\to K^+\pi^0\eta$}},
   { JHEP} {\bf 09} (2022) 107.

\bibitem{BESIII:2022ewq}
M.~Ablikim \textit{et al.}, (BESIII Collaboration), { {Amplitude analysis and
  branching-fraction measurement of $D_{s}^{+} \to
  \pi^{+}\pi^{0}\eta^{\prime}$}},  { JHEP} {\bf 04} (2022) 058.

\bibitem{Zhou:2015jba}
S.-H. Zhou, Y.-B. Wei, Q.~Qin, Y.~Li, F.-S. Yu, and C.-D. Lu, { {Analysis of
  two-body charmed $B$ meson decays in factorization-assisted
  topological-amplitude approach}},  { Phys. Rev. D} {\bf 92}, 094016 (2015).
 
\bibitem{Zhou:2016jkv}
S.-H. Zhou, Q.-A. Zhang, W.-R. Lyu, and C.-D. L\"u, { {Analysis of charmless
  two-body B decays in factorization-assisted topological-amplitude approach}},
   { Eur. Phys. J. C} {\bf 77}, 125 (2017).

\bibitem{Jiang:2017zwr}
H.-Y. Jiang, F.-S. Yu, Q.~Qin, H.-n. Li, and C.-D. L\"u, {
  {$D^0$-$\overline{D}^0$ mixing parameter $y$ in the factorization-assisted
  topological-amplitude approach}},  { Chin. Phys. C} {\bf 42}, 063101 (2018).

\bibitem{Wang:2017ksn}
D.~Wang, F.-S. Yu, P.-F. Guo, and H.-Y. Jiang, { {$K_{S}^{0}-K_{L}^{0}$
  asymmetries in $D$-meson decays}},  { Phys. Rev. D} {\bf 95}, 073007 (2017).

\bibitem{Zhou:2019crd}
S.-H. Zhou and C.-D. L\"u, { {Extraction of the CKM phase $\gamma$ from the
  charmless two-body $B$ meson decays}},  { Chin. Phys. C} {\bf 44}, 063101 (2020).
  
\bibitem{Zhou:2021yys}
S.-H. Zhou, R.-H. Li, Z.-Y. Wei, and C.-D. Lu, { {Analysis of three-body
  charmed B-meson decays under the factorization-assisted topological-amplitude
  approach}},  { Phys. Rev. D} {\bf 104}, 116012 (2021).
  
\bibitem{Zhou:2023lbc}
S.-H. Zhou, X.-X. Hai, R.-H. Li, and C.-D. Lu, { {Analysis of three-body
  charmless B-meson decays under the factorization-assisted
  topological-amplitude approach}},  { Phys. Rev. D} {\bf 107}, 116023 (2023).
  
\bibitem{Zhou:2024qmm}
S.-H. Zhou, R.-H. Li, and X.-Y. L\"u, { {Analysis of three-body decays
  B\textrightarrow{}D(V\textrightarrow{})PP under the factorization-assisted
  topological-amplitude approach}},  { Phys. Rev. D} {\bf 110}, 056001 (2024).
  
\bibitem{Zhou:2025}
S.-H. Zhou, \textit{et al.}, { {Updated analysis of $D \to PV$ under the
  factorization-assisted topological-amplitude approach}},
  in preparation.

\bibitem{Cheng:2024hdo}
H.-Y. Cheng and C.-W. Chiang, { {Updated analysis of
  $ D\to PP,VP$, and VV decays: Implications for $K_S^0-K_L^0$ asymmetries
  and $D^0-\bar {D^0}$ mixing}},  { Phys. Rev. D} {\bf 109}, 073008 (2024).
  
  \bibitem{Li:2018psm}
Y.~Li, W.-F. Wang, A.-J. Ma, and Z.-J. Xiao, {{Quasi-two-body decays
  $B_{(s)}\to K^*(892)h\to K\pi h$ in perturbative QCD approach}},  {Eur.
  Phys. J. C} {\bf 79} 37 (2019)
\bibitem{Ma:2019qlm}
A.-J. Ma, W.-F. Wang, Y.~Li, and Z.-J. Xiao, { {Quasi-two-body decays $B \to
  D K^*(892) \to D K \pi$ in the perturbative QCD approach}},  {Eur. Phys.
  J. C} {\bf 79} 539 (2019).

\bibitem{Fan:2020gvr}
Y.-Y. Fan and W.-F. Wang, { {Resonance contributions $\phi (1020,
  1680)\rightarrow K\bar{K}$ for the three-body decays $B\rightarrow K\bar{K}
  h$}},  { Eur. Phys. J. C} {\bf 80} 815 (2020).

\bibitem{Ma:2020jsb}
A.-J. Ma and W.-F. Wang, { {Contributions of the kaon pair from $\rho(770)$
  for the three-body decays $B \to DK\bar K$}},  {Phys. Rev. D} {\bf 103}
 016002 (2021).

\bibitem{Wang:2020nel}
W.-F. Wang, { {Contributions for the kaon pair from $\rho(770)$,
  $\omega(782)$ and their excited states in the $B\to K\bar K h$ decays}},
  {Phys. Rev. D} {\bf 103} 056021 (2021)

\bibitem{Wang:2024enc}
W.-F. Wang, L.-F. Yang, A.-J. Ma, and A.~Ramos, {{Low-mass enhancement of
  kaon pairs in $B^+\to \bar D^{(*)0} K^+\bar K^0 $ and
 $ B^0 \to D^{(*)-}K^+\bar K^{0}$ decays}},  { Phys. Rev. D}
  {\bf 109} 116009 (2024).

\bibitem{Blatt:1952ije}
J.~M. Blatt and V.~F. Weisskopf, { {Theoretical nuclear physics}}.
\newblock Springer, New York, 1952.

\bibitem{Aaij:2014baa}
 R.~Aaij \textit{et al.}, (LHCb Collaboration), { {Dalitz plot analysis of $B_s^0
  \rightarrow \bar{D}^0 K^- \pi^+$ decays}},  {Phys. Rev. D} {\bf 90}, 072003
  (2014).

\bibitem{Aaij:2016fma}
R.~Aaij \textit{et al.}, (LHCb Collaboration), {{Amplitude analysis of $B^{-}
  \to D^{+} \pi^{-} \pi^{-}$ decays}},  { Phys. Rev. D} {\bf 94}, 072001 (2016).
  
\bibitem{LHCb:2019sus}
R.~Aaij \textit{et al.}, (LHCb Collaboration), { {Amplitude analysis of the $B^+
  \rightarrow \pi^+\pi^+\pi^-$ decay}},  { Phys. Rev. D} {\bf 101}, 012006 (2020).
  
\bibitem{ParticleDataGroup:2024cfk}
S.~Navas \textit{et al.}, (Particle Data Group Collaboration), { {Review of
  particle physics}},  { Phys. Rev. D} {\bf 110}, 030001 (2024).

\bibitem{Cheng:2013dua}
H.-Y. Cheng and C.-K. Chua, {{Branching fractions and direct CP violation
  in charmless three-body decays of B mesons}},  { Phys. Rev. D} {\bf 88}, 114014
  (2013).

\bibitem{Bruch:2004py}
C.~Bruch, A.~Khodjamirian, and J.~H. Kuhn, { {Modeling the pion and kaon
  form factors in the timelike region}},  { Eur. Phys. J. C} {\bf 39}, 41 (2005).

\bibitem{CLEO:2009svp}
D.~Besson \textit{et al.}, (CLEO Collaboration), { {Improved measurements of D
  meson semileptonic decays to pi and K mesons}},  { Phys. Rev. D} {\bf 80}, 032005
  (2009).

\bibitem{Belle:2006idb}
 L.~Widhalm \textit{et al.},  (Belle Collaboration), { {Measurement of $D^0 \to \pi \ell \nu (K \ell \nu)$ 
 Form Factors and Absolute Branching
  Fractions}},  { Phys. Rev. Lett.} {\bf 97}, 061804 (2006).

\bibitem{BESIII:2018xre}
  M.~Ablikim \textit{et al.}, (BESIII Collaboration), { {First Measurement of the
  Form Factors in $D^+_{s}\rightarrow K^0 e^+\nu_e$ and $D^+_{s}\rightarrow
  K^{*0} e^+\nu_e$ Decays}},  { Phys. Rev. Lett.} {\bf 122}, 061801 (2019).

\bibitem{BESIII:2024zft}
 M.~Ablikim \textit{et al.}, (BESIII Collaboration), {{Improved measurement of
  the semileptonic decay $D_s^+ \to K^0 e^+ \nu_e$ }},  {
  Phys. Rev. D} {\bf 110}, 052012 (2024).

\bibitem{Bernard:2009ke}
C.~Bernard \textit{et al.}, { {Visualization of semileptonic form factors from
  lattice QCD}},  { Phys. Rev. D} {\bf 80}, 034026 (2009).

\bibitem{Du:2003ja}
D.-S. Du, J.-W. Li, and M.-Z. Yang, { {Form-factors and semileptonic decay
  of $D^+_s \to\phi\bar{l}\nu$ from QCD sum rule}},  {Eur. Phys. J. C} {\bf
  37}, 173 (2004).

\bibitem{Tian:2024lrn}
H.-J. Tian, Y.-L. Yang, D.-D. Hu, H.-B. Fu, T.~Zhong, and X.-G. Wu, {
  {Searching for $|V_{cd}|$ through the exclusive decay
  $D_s^+ \to K^0 e^+ \nu_e$ within QCD sum rules}},  { Phys.
  Lett. B} {\bf 857}, 138975 (2024).

\bibitem{Melikhov:2000yu}
D.~Melikhov and B.~Stech, {{Weak form-factors for heavy meson decays: An
  Update}},  {Phys. Rev. D} {\bf 62}, 014006 (2000).

\bibitem{Chen:2009qk}
C.-H. Chen, Y.-L. Shen, and W.~Wang, { { $|V_{(ub)}|$ and $B \to \eta^{(\prime)}$
Form Factors in Covariant Light Front Approach}},  { Phys. Lett. B}
  {\bf 686}, 118 (2010).

\bibitem{Verma:2011yw}
R.~C. Verma, { {Decay constants and form factors of s-wave and p-wave mesons
  in the covariant light-front quark model}},  { J. Phys. G} {\bf 39}, 025005 (2012).

\bibitem{Wang:2008ci}
W.~Wang and Y.-L. Shen, { {$D_s \to K, K^*, \phi$ form factors in
  the Covariant Light-Front Approach and Exclusive Ds Decays}},  { Phys.
  Rev. D} {\bf 78}, 054002 (2008).

\bibitem{Hu:2024tmc}
D.-D. Hu, X.-G. Wu, L.~Zeng, H.-B. Fu, and T.~Zhong, { {Improved light-cone
  harmonic oscillator model for the \ensuremath{\phi}-meson longitudinal
  leading-twist light-cone distribution amplitude and its effects to
  $D_s \to \phi \ell^+ \nu_\ell$}},
  { Phys. Rev. D} {\bf 110}, 056017 (2024).

\bibitem{Fajfer:2005ug}
S.~Fajfer and J.~F. Kamenik, { {Charm meson resonances and $D \to V$ 
semileptonic form-factors}},  { Phys. Rev. D} {\bf 72}, 034029
  (2005).

\bibitem{Momeni:2022gqb}
S.~Momeni and M.~Saghebfar, { {Semileptonic $D$ meson decays to the vector,
  axial vector and scalar mesons in Hard-Wall AdS/QCD correspondence}},  {
  Eur. Phys. J. C} {\bf 82}, 473 (2022).

\bibitem{Ahmed:2023pod}
H.~A. Ahmed, Y.~Chen, and M.~Huang, { {$D_{(s)}$- mesons semileptonic form
  factors in four-flavor holographic QCD}},  { Phys. Rev. D} {\bf 109}, 026008
  (2024).
  
  \bibitem{Fu-Sheng:2011fji}
Y.~Fu-Sheng, X.-X. Wang, and C.-D. Lu, { {Nonleptonic two-body decays of
  charmed mesons}},  {Phys. Rev. D} {\bf 84}, 074019 (2011).

\bibitem{BaBar:2007dro}
B.~Aubert \textit{et al.},  (BaBar Collaboration), { {Measurement of CP Violation
  Parameters with a Dalitz Plot Analysis of $B^\pm \to D(\pi^+ \pi^{-} \pi^0$ )
  $K^\pm$}},  { Phys. Rev. Lett.} {\bf 99}, 251801 (2007).
  
\bibitem{Belle:2006wbx}
A.~Kuzmin \textit{et al.}, (Belle Collaboration), { {Study of $\bar {B^0} \to D^0 \pi^+ \pi^-$ decays}},  { Phys. Rev. D} {\bf 76}, 012006 (2007).
 
\bibitem{LHCb:2016lxy}
 R.~Aaij \textit{et al.}, (LHCb Collaboration), { {Amplitude analysis of $B^{-}
  \to D^{+} \pi^{-} \pi^{-}$ decays}},  {Phys. Rev. D} {\bf 94}, 072001 (2016).

\end{thebibliography}\endgroup

\end{document}